\documentclass[amssymb,nobibnotes,superscriptaddress,longbibliography,notitlepage,twocolumn]{revtex4-1}

\usepackage{enumerate,amsmath,amsthm}
\usepackage{graphicx}
\usepackage{color}
\usepackage{comment}
\usepackage{braket}
\usepackage{bm}
\usepackage{caption}
\usepackage{subcaption}
\usepackage{url}

\usepackage{algorithm}
\usepackage{algpseudocode}

\usepackage{mathtools}

\allowdisplaybreaks

\begin{document}

\title{Comparative study of decoding the surface code using simulated annealing under depolarizing noise}
\author{Yusaku Takeuchi}
\email{u256654e@ecs.osaka-u.ac.jp}
\author{Yugo Takada}
\affiliation{Graduate School of Engineering Science, Osaka University, \\1-3 Machikaneyama, Toyonaka, Osaka, 560-8531, Japan}

\author{Tatsuya Sakashita}
\affiliation{Center for Quantum Information and Quantum Biology, Osaka University, Osaka, 560-0043, Japan}

\author{Jun Fujisaki}
\author{Hirotaka Oshima}
\author{Shintaro Sato}
\affiliation{Quantum Laboratory, Fujitsu Research, Fujitsu Limited, \\4-1-1 Kawasaki, Kanagawa 211-8588, Japan}
\affiliation{Fujitsu Quantum Computing Joint Research Division, Center for Quantum Information and Quantum Biology, Osaka University, \\1-2 Machikaneyama, Toyonaka, Osaka, 565-8531, Japan}

\author{Keisuke Fujii}
\email{fujii@qc.ee.es.osaka-u.ac.jp}
\affiliation{Graduate School of Engineering Science, Osaka University, \\1-3 Machikaneyama, Toyonaka, Osaka, 560-8531, Japan}
\affiliation{Fujitsu Quantum Computing Joint Research Division, Center for Quantum Information and Quantum Biology, Osaka University, \\1-2 Machikaneyama, Toyonaka, Osaka, 565-8531, Japan}
\affiliation{Center for Quantum Information and Quantum Biology, Osaka University, 560-0043, Japan}
\affiliation{RIKEN Center for Quantum Computing (RQC), Wako Saitama 351-0198, Japan}

\date{\today}

\begin{abstract}
We explored decoding methods for the surface code under depolarizing noise by mapping the problem into the Ising model optimization. 
We consider two kinds of mapping with and without a soft constraint and also various optimization solvers, including simulated annealing implemented on a CPU, 
"Fujitsu Digital Annealer" (DA), a hardware architecture specialized for the Ising problems,
and CPLEX, an exact integer programming solver. 
We find that the proposed Ising-based decoding approaches provide higher accuracy compared to the minimum-weight perfect matching (MWPM) algorithm for depolarizing noise
and comparable to minimum distance decoding using CPLEX.
While decoding time is longer than MWPM when we compare it with a single core CPU, our method is amenable to parallelization and easy to implement on dedicated hardware, suggesting potential future speedups. 
Regarding the mapping methods to the Ising model with and without a soft constraint, the SA decoder yielded higher accuracy without a soft constraint. 
In contrast, the DA decoder shows less difference between the two mapping methods,
which indicates that DA can find a better solution with smaller number of iterations even under the soft constraint.
Our results are important for devising efficient and fast decoders feasible with quantum computer control devices.
\end{abstract}

\maketitle

\section{Introduction}
To realize the theoretically proven computational speedup by quantum computers, 
fault-tolerant quantum computation using quantum error correction (QEC) is essential~\cite{fujii_topological_code}. Quantum computers on the scale of 100 qubits are becoming a reality~\cite{google_supremacy,IBM_utility}, and demonstration experiments of error correction~\cite{google_qec}, the most crucial building block of fault-tolerant quantum computers are underway.
Depending on the physical error probability and the error correction code used, 
many physical qubits are required to protect a single logical qubit from errors, 
and the overhead of fault-tolerant quantum computers becomes enormous~\cite{2048_factoring,Yoshioka_Mizukami}. 
The most important and challenging element to reduce such overhead is 
to decode at high speed and high accuracy, 
reducing the logical error probability as much as possible. 
For the surface codes, one of the most promising candidates
for the experimental realization of QEC, 
there is a known efficient decoding algorithm using minimum-weight perfect matching (MWPM)~\cite{MWPM,surface_Fowler} 
that efficiently performs minimum distance decoding for independent $X$ (bit flip) and $Z$ (phase flip) errors. 
However, for cases where $X$, $Y$, and $Z$ errors occur, such as depolarizing noise, MWPM is not optimal, because of the presence of $Y$ errors. 
This is one of the reasons why decoders beyond the MWPM algorithm have been explored such as using neural networks~\cite{Suzuki} and tensor network~\cite{tensornet_decoder}.
Moreover, for the color codes~\cite{color_code,takada} 
and even more general quantum low-density parity-check codes~\cite{Delfosse,LDPC_IBM,LDPC_Lukin}, 
performing decoding with high accuracy and efficiency is a challenging problem.

So far, by utilizing the mathematical correspondence between quantum error correction and the classical Ising model~\cite{fujii_topological_code}, the decoding problem of quantum error correction codes has been mapped to the optimization problem of the Ising model~\cite{fujii_prx,fujisaki_1,fujisaki_2,takada}, and decoding using heuristic solvers of the Ising model has been considered. 
There are two main approaches. One is to map the error distribution to Ising spins, impose syndrome constraints as soft constraints, and find the minimum error configuration that satisfies the constraints~\cite{fujii_prx,fujisaki_1,fujisaki_2}. 
Another approach is to map stabilizer operators 
to Ising spins and find the minimum error configuration 
where the syndrome constraint is automatically satisfied~\cite{takada}. 
It is not well understood which of these two approaches performs better 
under what circumstances. 

In this work, 
we investigate decoding methods based on optimization of the Ising model 
using various optimization solvers and evaluate their performance 
to determine which mapping, with or without constraints, is advantageous. 
Specifically, we consider the surface code under the depolarizing noise, 
where minimum distance decoding becomes harder.
As optimization problem solvers, 
we compared decoding accuracy and time using simulated annealing executed on a conventional CPU (using open-source software OpenJij~\cite{openjij}), 
``Fujitsu Digital Annealer" (DA), a hardware architecture
designed for Ising problems~\cite{da1,da2,da3,da4}, 
and CPLEX that can solve integer programming problems exactly~\cite{cplex}. 
First, regarding the decoding performance, 
it will be found that 
Ising-based decoders provide higher accuracy than MWPM for depolarizing noise and are comparable to that of the exact minimum distance decoding. 
On the other hand, in terms of decoding time, 
it takes longer time than MWPM when using a single CPU core.
However, the proposed approach is easy to parallelize.
Also it is easy to implement on a dedicated hardware such as field programable gate array (FPGA), so further acceleration is expected in the future. 

Regarding the mapping methods with and without constraints to the Ising model, 
higher decoding accuracy is obtained for the mapping without the constraint in the case of SA using OpenJij. 
This is probably because in SA  using OpenJij, satisfying the soft constraint is prioritized, and it gets trapped in sub-optimal solutions. 
On the other hand, in the case of using DA, 
there is almost no difference between the mappings with and without the constraint. 
This is because DA uses 
replica-exchange Monte Carlo
and is designed to be advantageous for energy landscapes with the soft constraint.
Regarding the decoding time, the number of steps required to find a solution is fewer for the method with the constraint. 
This is because 
when errors are mapped to Ising spins (with the soft constraint), 
in regions where the error probability is small, 
it is possible to find an optimal solution that locally satisfies the constraints in a small number of steps. 
On the other hand, when stabilizer operators are mapped to Ising spins (without the constraint), 
even in the regions with a small error probability, 
a sufficiently long relaxation time is required to decode. 
This knowledge is essential for constructing a high-performance, high-speed decoder that can be easily implemented with control devices close to quantum computers.

The rest of the paper is organized as follows.
In Sec \ref{surface_code}, we provide a preliminary explanation of the surface code and its decoding. 
In Sec \ref{Formulating_decoding_problem_via_Ising_model}, we provide the Ising-based formulations of the decoding problem
with and without the soft constraint.
In Sec \ref{Numerical_Simulation}, we provide details of the numerical simulations performed.
Section \ref{Conclusion_and_Discussion} is devoted to the conclusion and discussion.

 \section{Surface code}

 \label{surface_code}
\begin{figure}[t]
    \centering{
        \includegraphics[width=8cm]{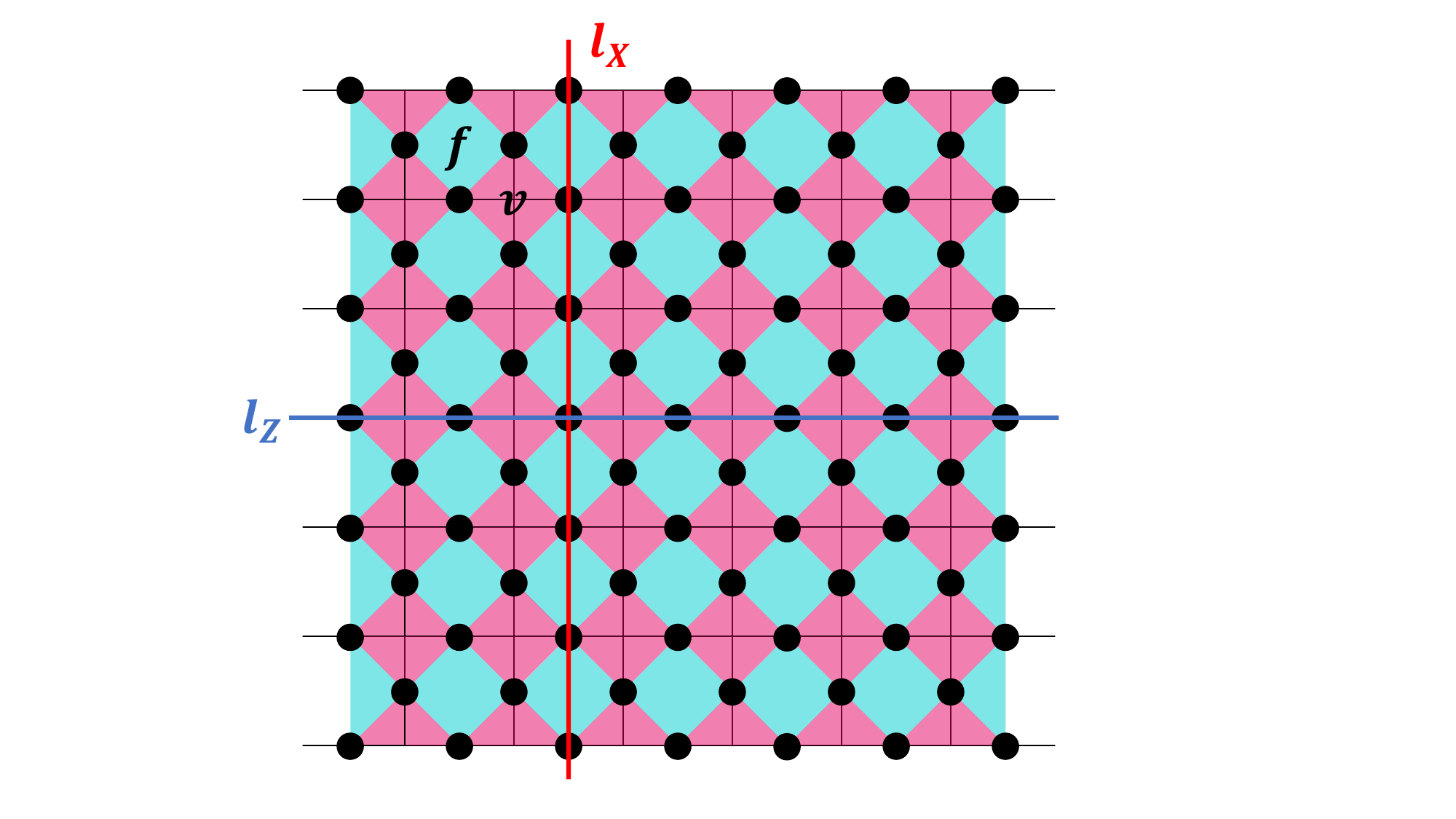}
        }
        \caption{The surface code ($d=7$). Black circles represent data qubits. 
        A $X$-type stabilizer operator $B_v$ is defined on the qubits located at the vertices of each red square. A $Z$-type stabilizer operator $A_f$ is defined on the qubits at the vertices of each blue square. 
        The product of $X$s on $l_x$ and the product of $Z$s on $l_z$ are the logical operators.}
        \label{surfacecode_def}
            
\end{figure}

The surface code \cite{surface1,surface2} is a type of stabilizer code, 
where a data qubit is aligned on each edge of a square lattice as shown in Fig. \ref{surfacecode_def}. 
The total number of qubits is $N_d:= d^2+(d-1)^2 $ in terms of 
the code distance $d$, the linear length of the square lattice.
The stabilizer generators are defined for each face $f$ and vertex $v$ by 
\begin{align}
    A_f&:=\prod_{i\in\partial f}Z_i, \;\;\; 
    B_v:=\prod_{i\in\delta v}X_i.
    \label{surfacecode_stabilizer}
\end{align}
Here, $Z_i$ and $X_i$ are the Pauli operators acting on the $i$th qubit. 
$\partial f$ represents the set of qubits on the edges surrounding face $f$, 
and $\delta v$ denotes the set of qubits connected to vertex $v$. 
As illustrated in Fig. \ref{surfacecode_def}, 
by using chains $l_x$ and $l_z$, each of which connects two opposite boundaries vertically and horizontally respectively,
the logical operators are defined as
\begin{align}
  L_X&:=\prod_{i\in l_{X}}X_i,  \;\;\;
  L_Z:=\prod_{j\in l_{Z}}Z_j.
  \label{stabilizer_surfacecode}
\end{align}

\begin{figure}[t]
    \centering{
        \includegraphics[width=8cm]{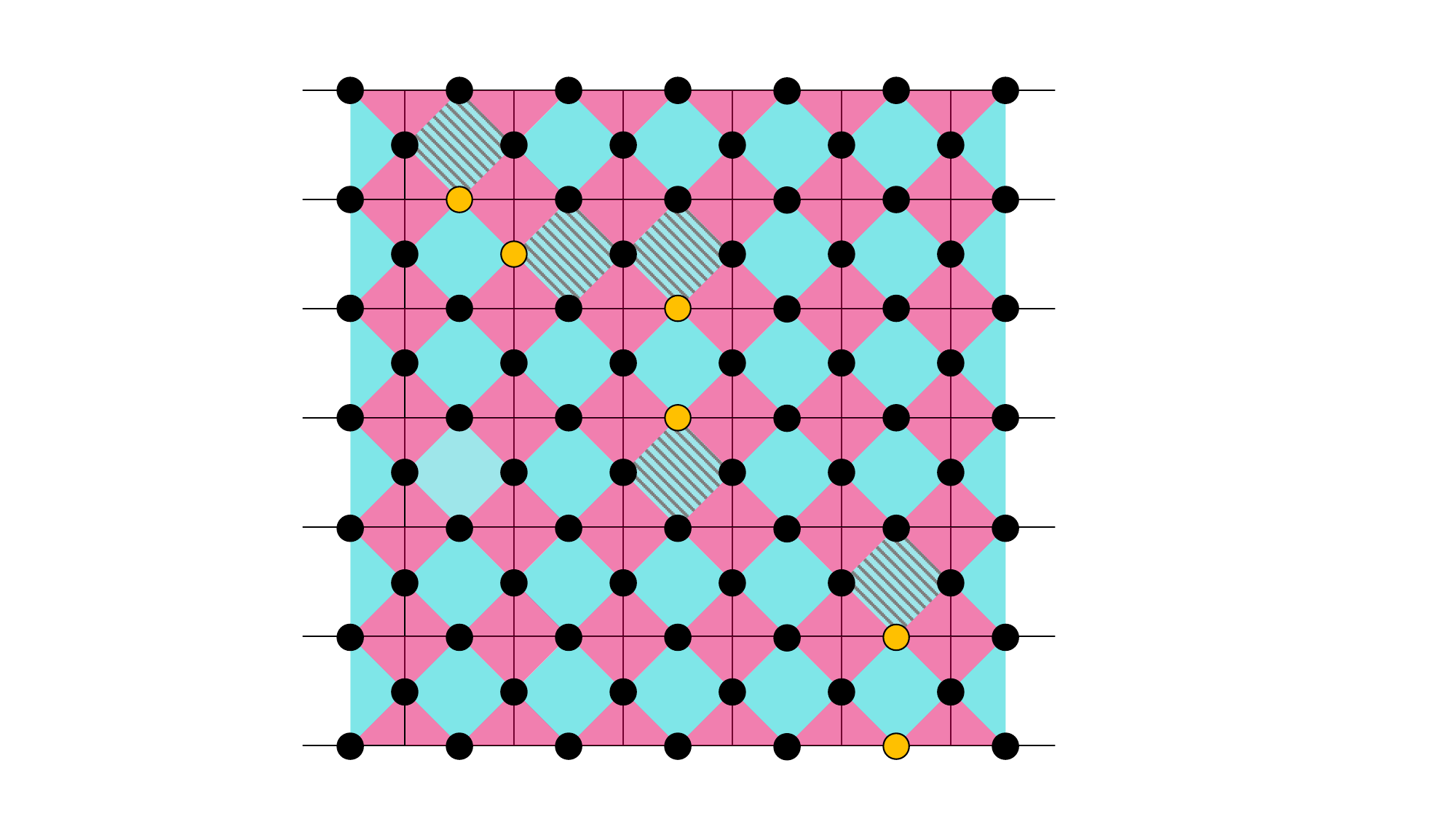}
        }
        \caption{Example of syndrome measurement of the surface code ($d=7$). Yellow circles indicate the data qubits affected by the Pauli $X$ errors. Shaded squares indicate stabilizer operators with their syndrome value $-1$.}
        \label{surfacecode_syndromemeas}
        
\end{figure}

Fig. \ref{surfacecode_syndromemeas} provides an example of the syndrome measurement. 
The stabilizer operators $A_f$ and $B_v$ are measured for each face $f$ and $v$ 
to detect $X$ and $Z$ errors, respectively.
Specifically, $Z$ errors being placed on edges form an error chain  and
the syndrome values are flipped at the boundary of the error chain.
Therefore, the location of $Z$ errors can be estimated by 
finding the path that connects the pairs of vertices with syndrome values $-1$,
which can be efficiently done by using minimum-weight perfect matching (MWPM) algorithm.
The same argument can be done for $X$ errors on the dual square lattice.
This type of decoding strategy can find an error of the highest probability conditioned on 
the observed syndrome values as long as $Z$ and $X$ errors occur independently.
However, in the presence of the $Y$ error, such as the depolarizing noise, 
$Z$ and $X$ errors occur in a correlated way, and hence 
the MWPM decoder for each of $X$ and $Z$ errors is far suboptimal.
The motivation for our research is to construct decoders that can more optimally correct such $Y$ errors as well.

\section{Formulating decoding problem via Ising model}
\label{Formulating_decoding_problem_via_Ising_model}
In this work, we aim to construct a decoder 
that can decode with high accuracy even in the presence of the $Y$ errors, such as depolarizing noise, 
by finding the error configuration with the highest probability conditioned on the observed syndrome values. 
For this purpose, we map the problem of finding the most likely error 
to the problem of finding the ground state of the Ising model.

There have been two different mappings; 
one assigns an Ising spin variable for each qubit (error)~\cite{fujii2017,fujisaki_1,fujisaki_2}, and 
the other for each stabilizer~\cite{takada}.
In the former case, the lowest energy spin configuration directly 
tells the most likely error location, while
the syndrome condition has to be imposed as a soft constraint.
In the latter case, while the lowest energy configuration itself is not 
related to the error configuration, instead the lowest energy corresponds to the minimum number of errors.
Then we can decide which of the logically different recovery operations 
minimizes the number of errors by comparing the minimum energy values among logically different recovery operations.

\subsection{Existing approach with constrained optimization}
Let us first explain about the mapping to the Ising Hamiltonian
with a soft constraint~\cite{fujii_prx,fujisaki_1,fujisaki_2}.
Suppose the Pauli $X$ error occurs on each data qubit independently with a probability $p$.
Representing this error with Ising spins, an error on the $i$th data qubit corresponds to a spin flip of $\sigma_i$ from $+1$ to $-1$.
Using this mapping, 
the Ising model Hamiltonian to find 
the error location from the syndrome values is defined as follows~\cite{fujii2017,fujisaki_1}:
\begin{equation}
    H=-J\sum_{f}^{N_f}a_f\prod_{i\in \partial f}^4\sigma_i-h\sum_{i}^{N_d}\sigma_i,
    \label{ising_hamiltonian_syndrome_constrain}
\end{equation}
where, $J$ and $h$, are hyperparameters, and $a_f$ corresponds to the syndrome value of the Pauli $Z$ stabilizer $A_f$.
$N_f$ and $N_d$ denote the number of Pauli $Z$ stabilizers and data qubits, respectively.
The first term imposes the syndrome condition, and 
the second term counts the number of errors.
Hence if $J$ is sufficiently large,
the ground state corresponds to the error location satisfying 
the syndrome condition with the smallest number of errors.

This approach introduces the soft constraint by imposing a penalty in the first term to find the arrangement of errors that satisfies the syndrome constraint with the smallest number of errors.
Therefore, it is very tricky to choose the hyperparameter $J$ appropriately.
Furthermore, a suboptimal solution is not always guaranteed to satisfy the syndrome condition.
To handle this, we here explore another mapping to the Ising Hamiltonian
without the soft constraint as explained below.

\subsection{Pauli error decomposition}
For a given syndrome $S$, any Pauli error $E\in\{I,X,Y,Z\}^{\otimes n}$ that satisfies the syndrome $S$ can be decomposed into 
a product of pure error $T(S)$, logical operator $L \in \mathcal{L}$, and stabilizer operator $G \in \mathcal{G}$ as follows~\cite{Poulin}:
\begin{equation}
\label{error_decomposition}
  E=T(S)LG, \quad L\in\mathcal{L},\quad G\in\mathcal{G}.
\end{equation}
The pure error $T(S)$ is the operator that returns the erroneous state to the original code space.
Although the construction of $T(S)$ is arbitrary, 
we here construct it by connecting error chains from the stabilizer operators with $-1$ syndrome to appropriate boundaries 
(e.g. top and bottom boundaries for $X$ error chains).

\subsection{Counting $X$ errors}
To find the most likely error, 
we have to count the total number of errors through the Ising model Hamiltonian.
Let us first consider the case where $E$ consists only of $X$ errors, 
and hence $G$ and $L$ are also products of $X$s.
We define a bitstring $e_k$ corresponding to the location of $X$ errors.
That is the $k$th bit value $e_k$ is 1 if the $k$th qubit has an $X$ error, and 0 otherwise.
The number of errors $N_e$ is counted by
\begin{equation}
\label{number_of_errors1}
  N_{e}=\sum_{k}^{N_d}e_k.
\end{equation}
Similarly to $e_k$, the bit strings for $T(S)$, $L$, and $G$ 
are defined as $t$, $l$, and $g$, respectively.
By rewriting Eq.~\eqref{error_decomposition} in terms of these binary variables,
we have
\begin{equation}
  \label{number_of_errors2}
  N_e=\sum_{k}^{N_d} ( t_k\oplus l_k\oplus g_k ).
\end{equation}
In order to obtain an Ising Hamiltonian,
we convert binary variable $b$ to spin variable $\bar{b}$ by
\begin{equation}
    \label{spin_binary_conversion}
    \bar{b}=1-2b.
\end{equation}
This provides 
\begin{align}
  N_e&=\sum_{k}^{N_d}\{(1-\bar{t}_k\bar{l}_k\bar{g}_k)/2\}\\
          &=\frac{N_d}{2}-\frac{1}{2}\sum_{k}^{N_d}(\bar{t}_k\bar{l}_k\bar{g}_k)\label{partition_function2}.
\end{align}
By omitting the constant, we can define an Ising Hamiltonian as follows:
\begin{equation}
  \label{energy_function2}
  H=-\sum_{k}^{N_d}\bar{t}_k\bar{l}_k\bar{g}_k.
\end{equation}
Since $\bar{t}_k$ is determined from a given syndrome $S$ and also $\bar{l}$ 
if we choose one possible logical operator $L$.
By replacing $\bar{t}_k\bar{l}_k$ with $J_k$
as a coupling constant defined by $S$ and $L$
we have
\begin{equation}
  \label{energy_function3}
  H=-\sum_{k}^{N_d}J_k\bar{g}_k.
\end{equation}
If the above Ising Hamiltonian is minimized 
with respect to $\{ \bar{l}_k \}$ and $\{ \bar{g}_k\}$, 
we can find the error configuration with a minimum number of errors.
However, the elements of $\{ \bar{g}_k\}$, i.e., stabilizer group, 
is exponentially many and does not take all possible Ising spin configurations, being subject to certain constraints.
In order to deal with this,
we introduce another Ising spin variable $\bar{\sigma}$ for each stabilizer generator,
where $+1$ and $-1$ indicate the stabilizer generator is included and not included respectively.
Let $\bar{\sigma}_a$ be the spin variable of face $a$ associated with the stabilizer $B_a$.
Then $\bar{g}_k$ is defined from $\bar{\sigma}_a$ as follows:
\begin{equation}
  \bar{g}_k=\bar{\sigma}_{l(k)}\bar{\sigma}_{r(k)},
    \label{eq:gauge_trans}
\end{equation}
where $l(k)$ and $r(k)$ is the indices for the pairs of stabilizer generators 
that contain $k$th qubit.
By doing this, the constraint on $\{\bar{g}_k\}$ is 
automatically satisfied with through $\{ \bar{\sigma}_a\}$.
Note that this is nothing but decomposing a stabilizer operator $G$ into a product of stabilizer generators.
Substituting Eq. (\ref{eq:gauge_trans}) into Eq.~\eqref{energy_function3},
we have
\begin{equation}
  H=-\sum_k J_k\bar{\sigma}_{l(k)}\bar{\sigma}_{r(k)}.
  \label{energy function4}
\end{equation}
Then 
\begin{align}
    \min _{ L, \{ \bar{\sigma} _k \} }  H
\end{align} 
tells us the minimum number of errors 
satisfying a given syndrome $S$, and hence the most likely error configuration.
Note that the syndrome condition is always satisfied no matter what configuration the spin variable $\sigma$ has.
Therefore, there is no need to add a penalty term to the Hamiltonian in order to impose the syndrome constraint compared to the previous approach.

\subsection{Counting $X$, $Y$, and $Z$ errors}
Under depolarizing noise, each data qubit can experience one of the Pauli errors, $X$, $Y$, and $Z$.
From the definition of the surface code Eq.~\eqref{stabilizer_surfacecode},
the Pauli $Z$ error can be treated similarly to the Pauli $X$ error
by considering the dual square lattice.
Let us define bits for the $k$th qubit with respect to the Pauli $X$ and $Z$ errors by $e^X_{k}$ and $e^Z_{k}$ respectively.

\begin{figure}[t]
    \centering{
        \includegraphics[width=8cm]{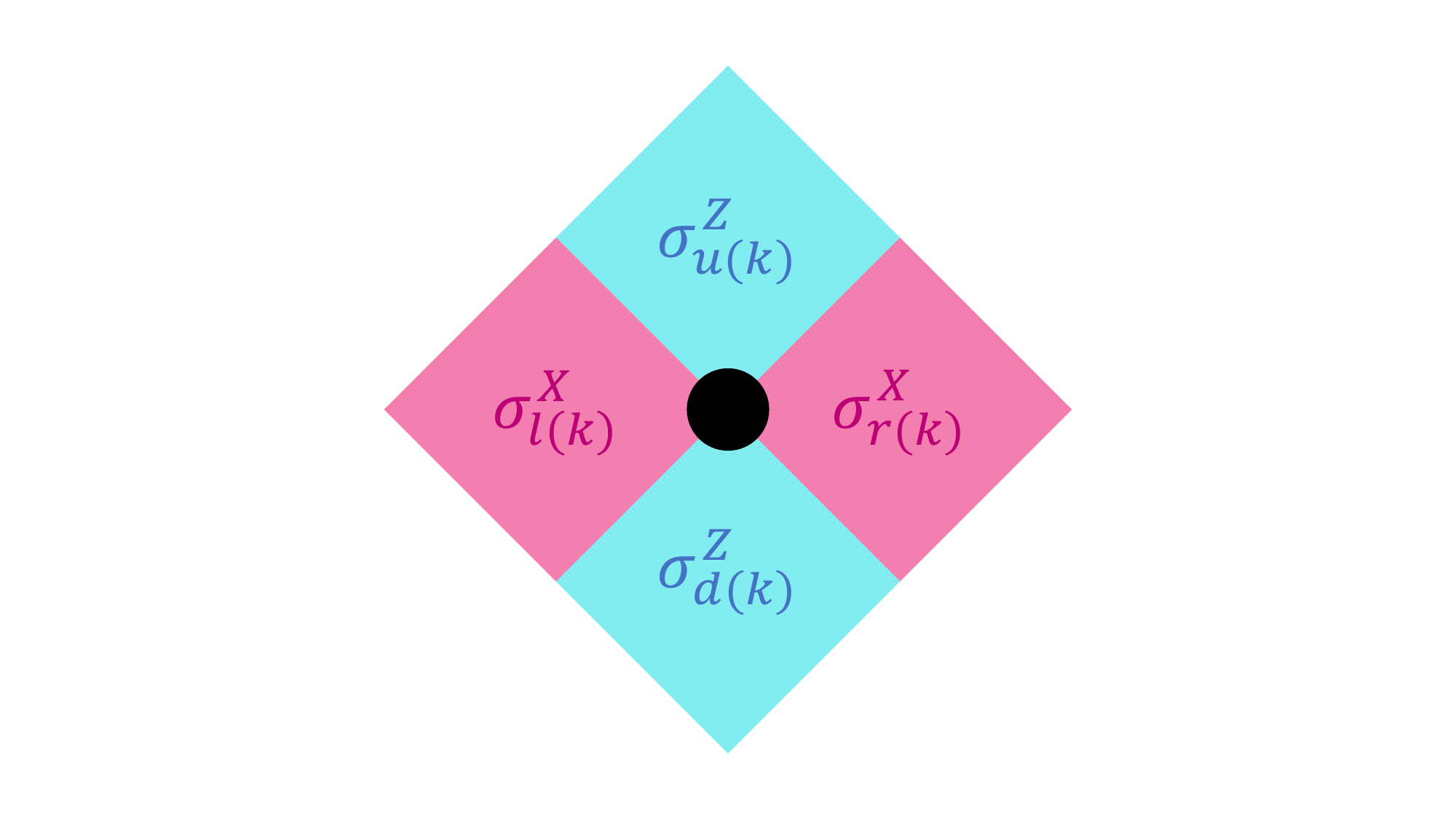}
        \caption{Ising spin variables and the surface code.
        Black circles represent the $k$th data qubits. The number of errors in the $k$th data qubit is counted in terms of the Ising spin variables corresponding to stabilizer operators neighboring to the data qubit. }
        \label{surfacecode_spin_value}
        }  
\end{figure}

Similarly to Eq. ~\eqref{number_of_errors1}, 
we have a total number of $X$ and $Z$ errors as follows:
\begin{equation}
    N_{e}=\sum_{k}^{N_d}e^X_k+\sum_{k}^{N_d}e^Z_k.
    \label{number_of_errors3}
\end{equation}
The $Y$ error is considered to be simultaneous $X$ and $Z$ errors,
i.e., $e^X_{k} = e^Z_{k} = 1$.
Therefore, Eq.~\eqref{number_of_errors3} doubly counts 
the number of the $Y$ errors.
Therefore we can count the total number of $X$, $Y$, and $Z$ errors
by subtracting the total number of $Y$ errors as follows:
\begin{equation}
    N_{e}=\sum_{k}^{N_d}e^X_k+\sum_{k}^{N_d}e^Z_k-\sum_{k}^{N_d}{e^X_ke^Z_k}.
    \label{number_of_errors4}
\end{equation}
By using the same argument with $X$ errors,
we can convert $N_e$ to Ising Hamiltonian
\begin{align}
    H=&-\sum_{k}^{N_d}J_{k}^X\sigma_{l(k)}^X\sigma_{r(k)}^X-\sum_{k}^{N_d}J_{k}^Z\sigma_{u(k)}^Z\sigma_{d(k)}^Z \nonumber \\
    &+\sum_{k}^{N_d}J_{k}^XJ_{k}^Z\sigma_{l(k)}^X\sigma_{r(k)}^X\sigma_{u(k)}^Z\sigma_{d(k)}^Z,
    \label{energy_function5}
\end{align}
where now we introduced another Ising spin variable for $Z$ errors
as shown in Fig.~\ref{surfacecode_spin_value}.

\subsection{Mapping to QUBO (two-body Ising model)}
As seen above, the Ising Hamiltonian
for counting $X$, $Y$, and $Z$ errors contains 
higher order terms, four-body Ising interactions.
On the other hand, 
various approximate solvers for optimization of the Ising models have been developed so far, ranging from software approaches running on conventional CPUs, 
i.e., simulated annealing (SA), 
to classical dedicated hardware approaches, such as 
"Fujitsu Digital Annealer" (DA)
to quantum annealing using a quantum device. 
Many of these approximate solvers solve the problem in the QUBO form~\cite{fujisaki_1}, 
which includes up to two-body interactions. 
Therefore, in order to take advantage of the assets of these solvers, 
we will map the four-body Ising interactions to the QUBO form by converting it to two-body interactions without changing the lowest energy configuration 
in the following.

Let us explain the reduction using binary valuable $b \in \{ 0,1\}$.
The product of binary variables $b_ib_j$ in high-degree terms is represented by the auxiliary binary variable $b'$:
\begin{equation}
    b_ib_j=b'.
    \label{depressing}
\end{equation}
Then by using the new variable $b'$ instead of $b_i b_j$, we 
can reduce the order of the term.
In order to impose the constrain Eq.~\eqref{depressing},
we have to add a penalty term $H_{\mathrm{penalty}}$:
\begin{equation}
\label{penalty_term_depressing}
    H_{\mathrm{penalty}}=\alpha(b_ib_j-2b_jb'-2b_ib'+3b'),
\end{equation}
where
$\alpha$ is a hyperparameter determining the penalty term's magnitude and is set by the user.
$H_{\mathrm{penalty}}$ only takes a positive value when Eq.~\eqref{depressing} isn't satisfied and is 0 when it is.

By using the above argument for the Ising Hamiltonian in Eq.~\eqref{energy_function5},
we have
\begin{align}
    H=&-\sum_{k}^{N_d}J_{k}^XH_{1(k)}-\sum_{k}^{N_d}J_{k}^ZH_{2(k)}+\sum_{k}^{N_d}J_{k}^XJ_{k}^ZH_{3(k)} \nonumber \\
      &+\sum_{k}^{N_d}H_{\mathrm{penalty}(k)}^X+\sum_{k}^{N_d}H_{\mathrm{penalty}(k)}^Z,
    \label{energy function6}
\end{align}
where
\begin{align}
    H_{1(k)} &=(1-2b_{l(k)}^X)(1-2b_{r(k)}^X),
    \\
    H_{2(k)}&=(1-2b_{u(k)}^Z)(1-2b_{d(k)}^Z),
    \\
    H_{3(k)}=& H_{1(k)} H_{2(k)} \nonumber \\
             =&1-2b_{l(k)}^X-2b_{r(k)}^X-2b_{u(k)}^Z-2b_{d(k)}^Z\nonumber \\
             &+4b_{l(k)}^Xb_{r(k)}^X+4b_{l(k)}^Xb_{u(k)}^Z+4b_{l(k)}^Xb_{d(k)}^Z \nonumber \\
             &+4b_{r(k)}^Xb_{u(k)}^Z+4b_{r(k)}^Xb_{d(k)}^Z+4b_{u(k)}^Zb_{d(k)}^Z \nonumber \\
             &-8b_{rl(k)}^Xb_{u(k)}^Z-8b_{rl(k)}^Xb_{d(k)}^Z-8b_{r(k)}^Xb_{ud(k)}^Z \nonumber\\
             &-8b_{l(k)}^Xb_{ud(k)}^Z+16b_{rl(k)}^Xb_{ud(k)}^Z,
\end{align}
and
\begin{align}
    H_{\mathrm{penalty}(k)}^X=&\alpha(b_{r(k)}^Xb_{l(k)}^X-2b_{l(k)}^Xb_{rl(k)}^X-2b_{r(k)}^Xb_{rl(k)}^X \nonumber \\
                         &+3b_{rl(k)}^X),
                         \\
    H_{\mathrm{penalty}(k)}^Z=&\alpha(b_{u(k)}^Zb_{d(k)}^Z-2b_{d(k)}^Zb_{ud(k)}^Z-2b_{u(k)}^Zb_{ud(k)}^Z \nonumber \\
                         &+3b_{ud(k)}^Z).
\end{align}

\section{Numerical Simulation}
\label{Numerical_Simulation}
To evaluate the performance of the proposed method, we perform a Monte Carlo simulation.
Depolarizing error with probability $p$,
\begin{align}
\mathcal{E}(\rho) = (1-p)\rho+\frac{p}{3}(X\rho X +Z\rho Z+Y\rho Y),
\end{align}
is introduced on each data qubit.
The error syndrome is measured ideally and then used 
to construct the Ising Hamiltonian.
For optimization of the constructed Ising Hamiltonian, 
we used SA and DA and evaluated the performance of each.
These results are compared with
the MWPM decoder, an efficient decoder for the surface code but suboptimal in the presence of $Y$ errors, 
and the integer programming (IP) decoder by using CPLEX to find the minimum weight Pauli error with a given syndrome pattern exactly.
Furthermore, we will also compare two mappings to the Ising models with and without the soft constraint later.

\subsection{Performance of SA decoder}
First, we perform decoding using SA,
where use the open-source library OpenJij~\cite{openjij}.
The parameters of OpenJij adopted in this section are shown in Table \ref{parameter_OpnJij}.
We evaluated the logical error probability with code distances $d = 3,5,7,9$.
The number of samples is taken to be 92,000 for each physical error probability.
\begin{table}[t]
\caption{Annealing parameter set of OpenJij}
\label{parameter_OpnJij}
\centering
\begin{tabular}{lc}
\hline 
Parameter & Value \\
\hline
$J$ & 30 \\
$\alpha$ \; in Eq. \eqref{penalty_term_depressing} & 30 \\
\begin{tabular}{l}
Number of divisions of \\
inverse temperature
\end{tabular}&
 100 \\
\begin{tabular}{l}
Number of steps at \\
each inverse temperature 
\end{tabular}&
$N_d \times$ (Number of variables)\\
\begin{tabular}{l}
Initial value of\\
inverse temperature 
\end{tabular}&
0.0001 \\
\begin{tabular}{l}
Target value of \\
each inverse temperature 
\end{tabular}&
5 \\
Number of repetitions & 2 \\
\hline 
\end{tabular}
\end{table}

\begin{figure}[!t]
\includegraphics[width=8cm]{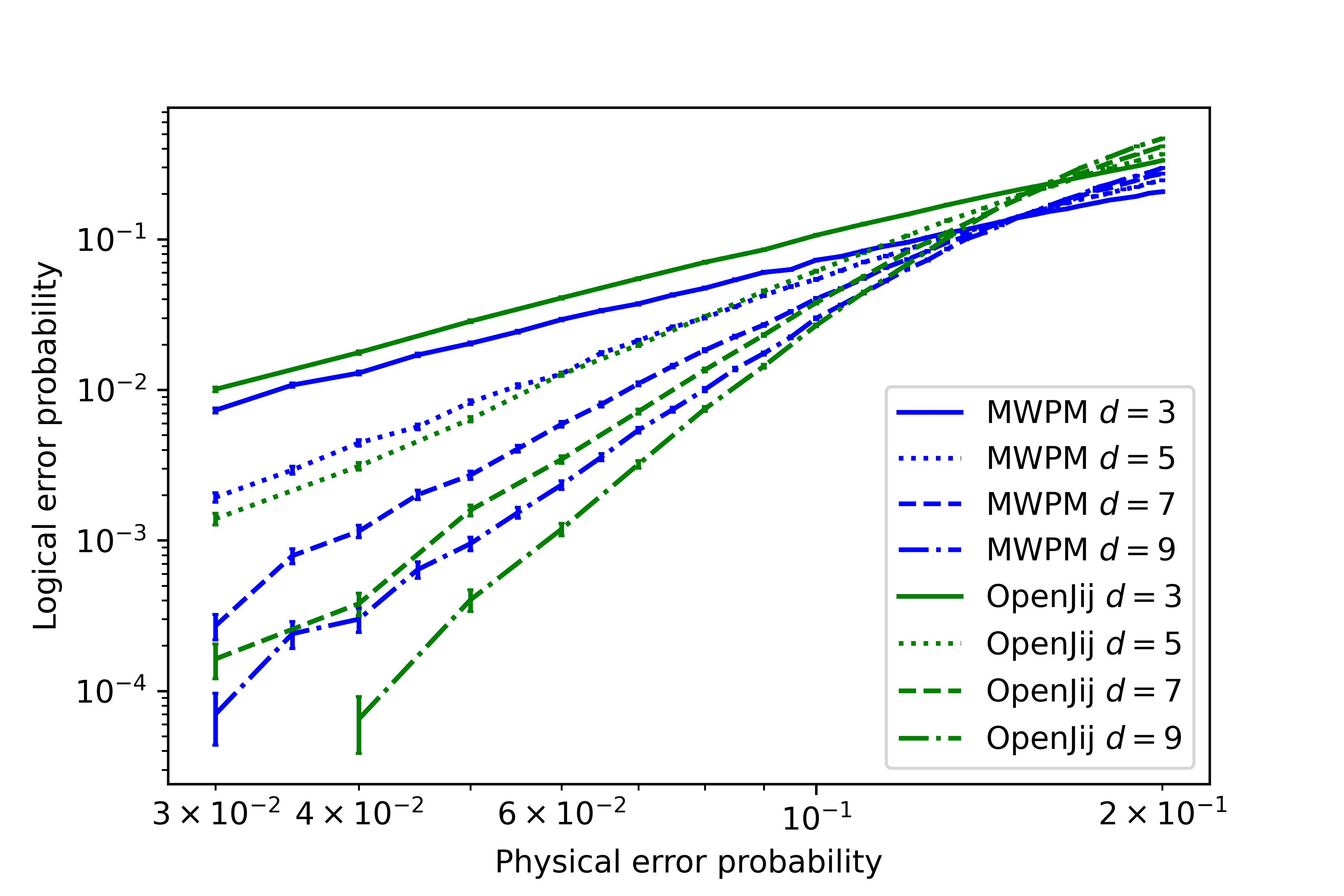}
\caption{The logical error probability is plotted as a function of the physical error probability with code distances $d=3,5,7,9$ for the SA(OpenJij) decoder (colored green) and
the MWPM decoder (colored blue).}
\label{surface_deporalizing_OpenJij_MWPM}
\end{figure}

\begin{figure}[!t]
\includegraphics[width=8cm]{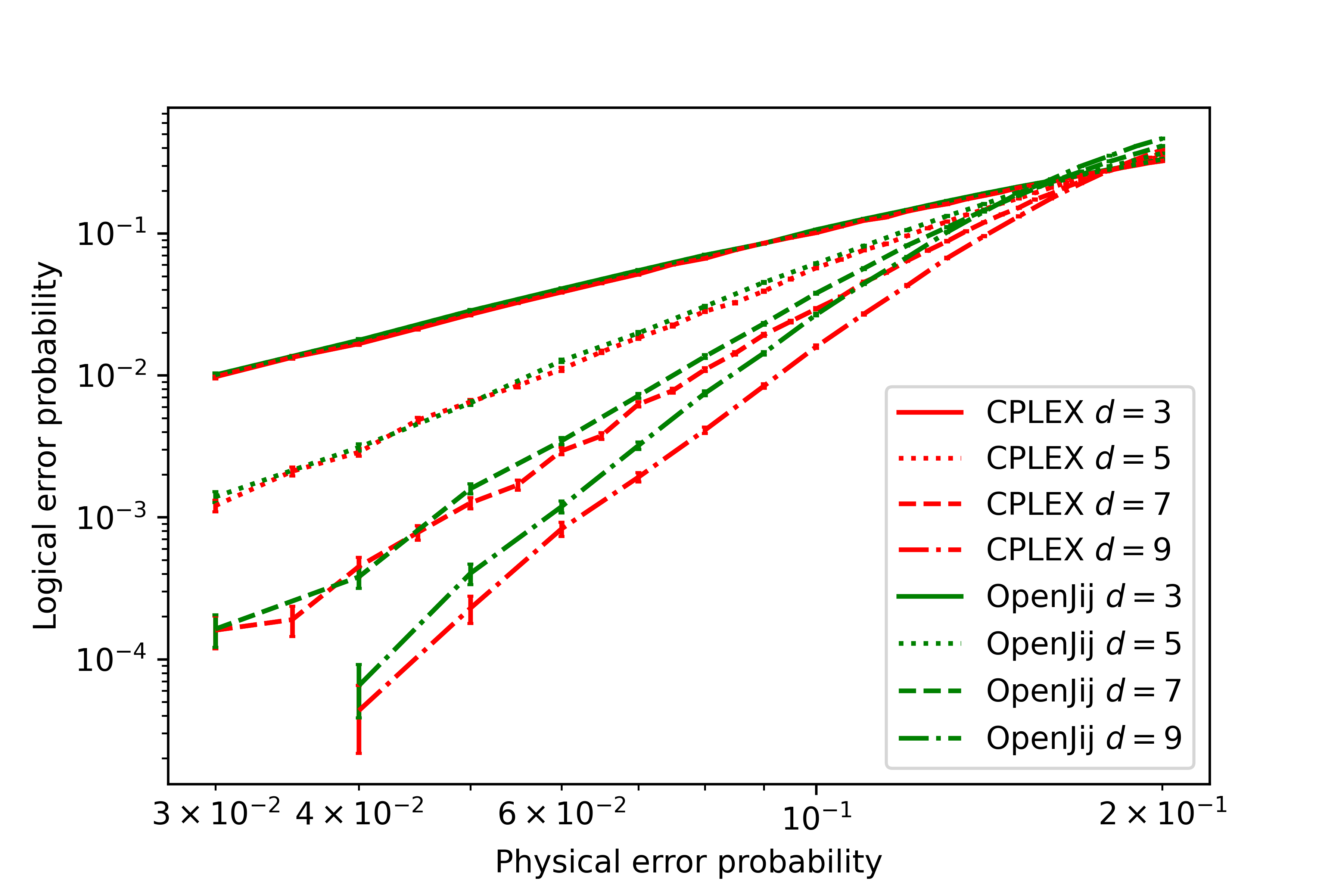}
\caption{The logical error probability is plotted as a function of the physical error probability with code distances $d=3,5,7,9$ for the SA(OpenJij) decoder (colored green) and
the CPLEX decoder (colored red).}
\label{surface_deporalizing_OpenJij_CPLEX}

\end{figure}

The logical error probabilities at each physical error probability are shown in Figs. \ref{surface_deporalizing_OpenJij_MWPM} and \ref{surface_deporalizing_OpenJij_CPLEX}.
From Fig.~\ref{surface_deporalizing_OpenJij_MWPM}, it can be seen that the logical error probability for the SA(OpenJij) decoder is lower than that of the MWPM decoder for code distances $d=5,7,9$.
In the case of $d=3$, because of the small size, 
the error patterns degenerate, and hence the performance depends on which solution was chosen by the decoding algorithm.
The threshold value of the physical error probability for SA(OpenJij) is between 16\% and 18\%, while that for MWPM is between 14\% and 16\%.
From these results, as we expected, the SA(OpenJij) decoder can decode with higher accuracy than the MWPM decoder.
We further compare the performance of the SA(OpenJij) decoder with the IP(CPLEX) decoder
in order to see how minimum weight Pauli error can be found in the SA(OpenJij) decoder.
From Fig.~\ref{surface_deporalizing_OpenJij_CPLEX}, it can be seen that SA(OpenJij) 
can decode with the same level of accuracy as IP(CPLEX).
Note that, 
since the logical error probabilities of both are the same for $d=3$, 
the minimum distance decoding is achieved in the case of $d=3$.

Now that we know that high performance can be obtained, let us next examine the time required for decoding.
At high error probabilities near the threshold, 
the syndrome flips a lot, which causes a lot of frustration and makes the optimization problem difficult. 
On the other hand, at low error probabilities, 
locally optimal solutions can be easily found. 
Therefore, we calculated and compared the average time of decoding for 100 samples for the two regions of high and low error probability.
The decoding time measured on a single CPU core shows that decoding time increases exponentially with IP(CPLEX) near the threshold, while SA(OpenJij) scales polynomially. 
The IP(CPLEX) decodes faster at low error probabilities for larger codes,
where the optimization problem becomes much easier.
However, these behaviors highly depend on annealing parameter settings. 
Although our parameters are chosen depending only on the code size, 
the decoding difficulty varies with error probability.
Optimizing annealing parameters for each error rate could reduce decoding time, while this is not our scope. 
Furthermore, while the SA(OpenJij) decoder is not as fast as the MWPM decoder,
SA is much simpler than MWPM and hence if it is implemented on dedicated hardware or massively parallel hardware,
we can further improve the decoding time.

\begin{figure}[!t]
  \begin{minipage}[b]{0.45\linewidth}
    \includegraphics[keepaspectratio, scale=0.25]{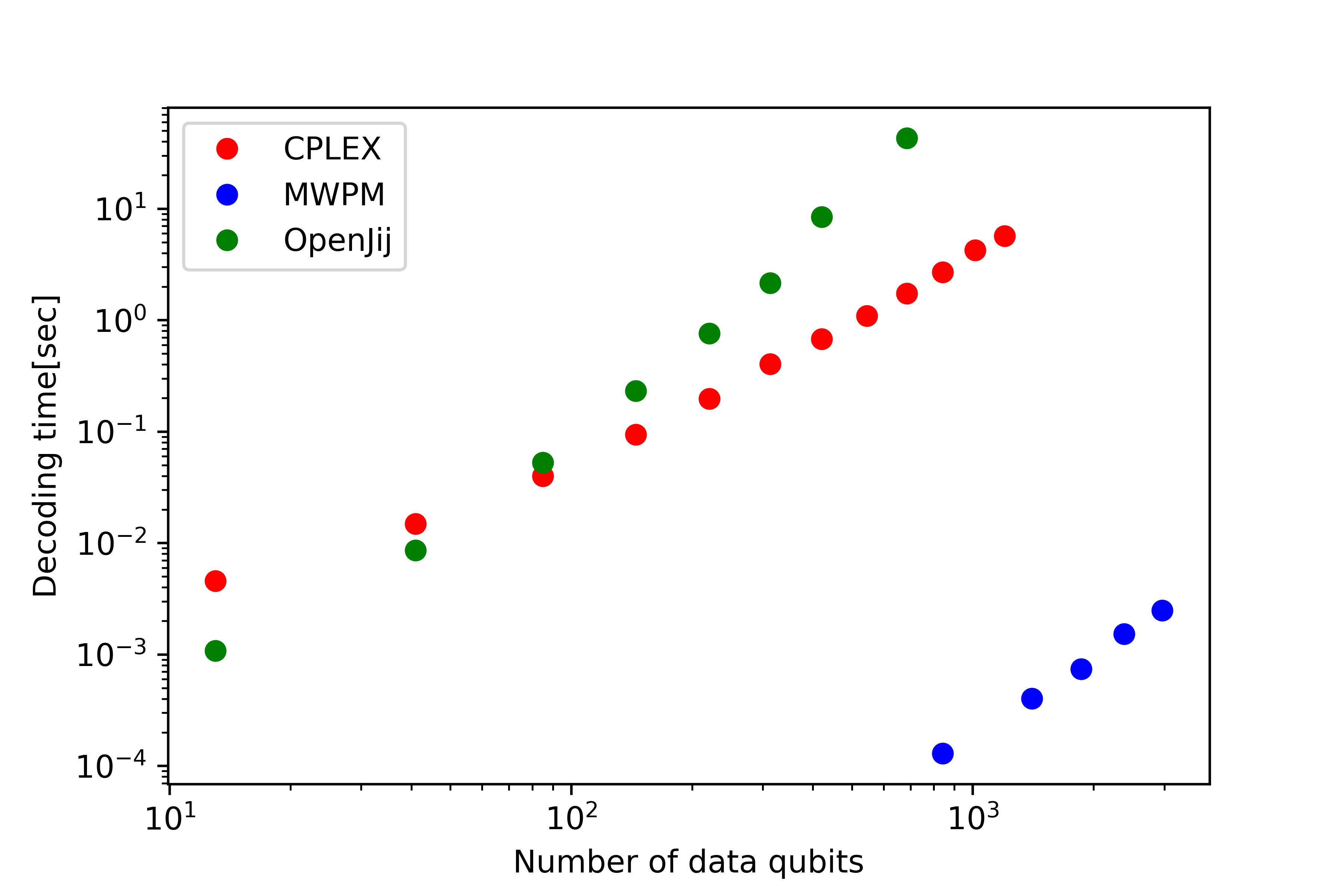}
    \subcaption{}
    \label{decoding_time_deporalizing5}
  \end{minipage}
  \begin{minipage}[b]{0.45\linewidth}
    \includegraphics[keepaspectratio, scale=0.25]{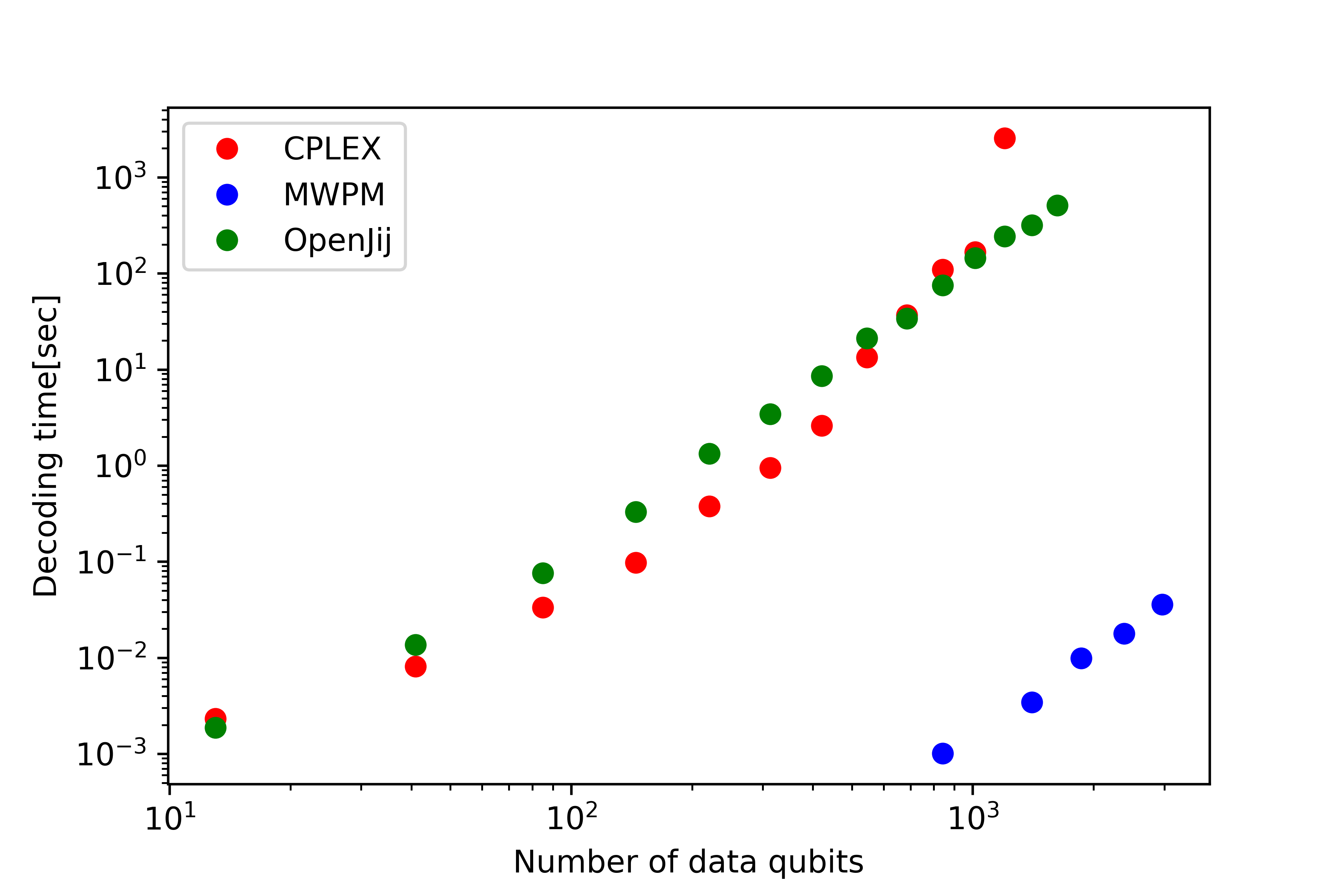}
    \subcaption{}
    \label{decoding_time_deporalizing15}
  \end{minipage}
  \caption{Decoding time as a function of the number of the qubits in the surface codes with physical error probabilities (a) $p=0.05$ and (b) $p=0.15$.}
  \label{decoding_time_deporalizing}
\end{figure}

\subsection{Performance of DA decoder}
\begin{figure}[!t]
        \includegraphics[width=8cm]{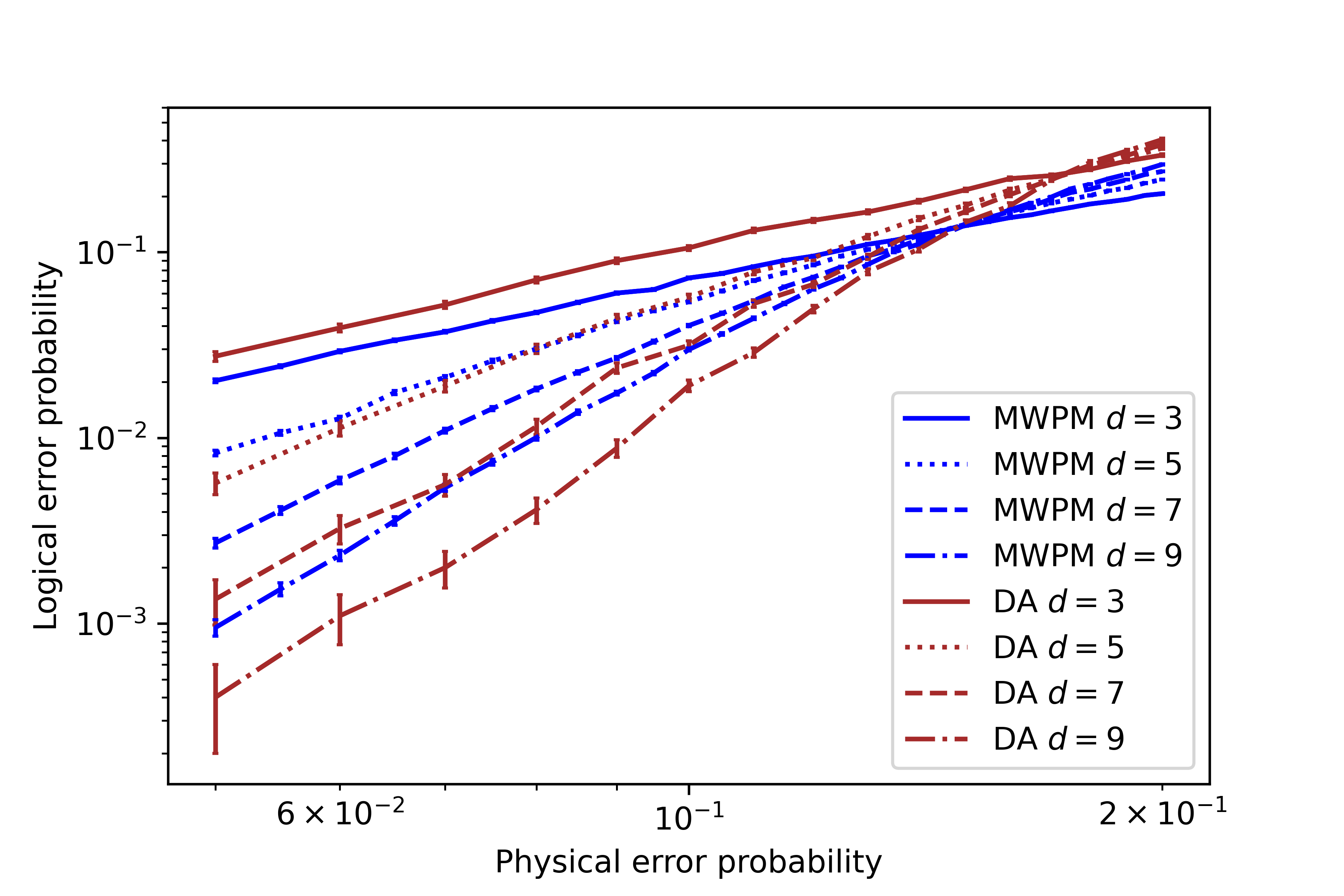}
        \caption{The logical error probability is plotted as a function of the physical error probability with code distances $d=3,5,7,9$ for the DA decoder (colored brown) and the MWPM decoder (colored blue).}
        \label{surface_deporalizing_DA_MWPM}
        
\end{figure}

\begin{figure}[!t]
    \centering{
        \includegraphics[width=8cm]{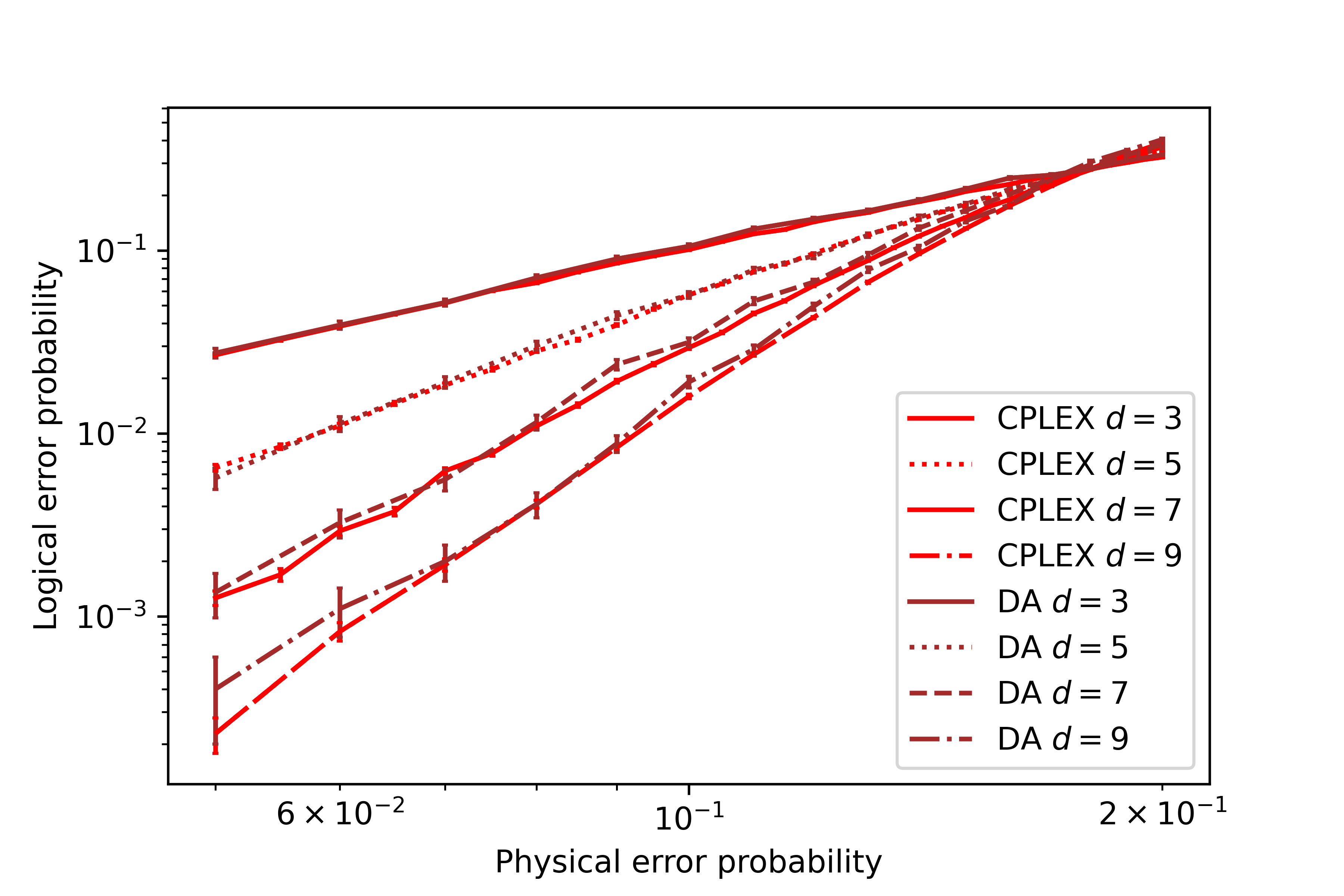}
        \caption{The logical error probability is plotted as a function of the physical error probability with code distances $d=3,5,7,9$ for the DA decoder (colored brown) and the CPLEX decoder (colored red).}
        \label{surface_deporalizing_DA_CPLEX}
        }  
\end{figure}

Next, we investigate the performance of the DA decoder,
where the Ising Hamiltonian is optimized by using a dedicated hardware designed to solve the Ising models.
The parameters of DA adopted in this section are shown in Table \ref{parameter_DA}.
Similar to the SA(OpenJij) decoder, 
we evaluated the logical error probability with code distances $d$ of 3, 5, 7, and 9.
We conducted simulations with 10,000 samples for each physical error probability and calculated the logical error probability.
The results are shown in Figs. \ref{surface_deporalizing_DA_MWPM} and \ref{surface_deporalizing_DA_CPLEX}.
As we expected, the DA decoder also exhibits higher accuracy than the MWPM decoder and is comparable to the IP(CPLEX) decoder.
Regarding the decoding time, 
as will be examined later, 
the mapping without the soft constraint requires sufficiently large number of iterations for convergence. Hence it takes almost the same decoding time $\sim 10^{-1}-10^{-2}$ sec when the number of qubits is $\sim 100$ as IP(CPLEX) and SA(OpenJij).

\begin{table}[!t]
    \caption{Annealing parameter set of DA}
    \label{parameter_DA}
      \centering
    \begin{tabular}{lc}
    \hline 
    Parameter                            &  Value \\
    \hline
        $J$                             &  1 \\
        $\alpha$(Equation \ref{penalty_term_depressing}) &  2.7 \\
        Annealing mode                  & Replica exchange \\
        Maximum temperature             & 5 \\
        Number of replicas              & 128  \\
        \hline 
    \end{tabular}
\end{table}


\subsection{Comparison between constrained or unconstrained optimization}

\begin{table}[b]
\caption{Annealing parameter set of OpenJij}
\label{parameter_OpenJij_phaseflip}
\centering
\begin{tabular}{lcc}
\hline 
Parameter & w & w/o \\
\hline
$J$ & 30 & 20\\
$h$ & --- & 20\\
$\alpha$(Equation \ref{penalty_term_depressing}) & ---  & 50 \\
\begin{tabular}{l}
Number of divisions of \\
inverse temperature
\end{tabular}&
\multicolumn{2}{c}{100}\\
\begin{tabular}{l}
Number of steps at \\
each inverse temperature 
\end{tabular}&
\multicolumn{2}{c}{$N_d$ $\times$ (Number of variables)}\\
\begin{tabular}{l}
Initial value of\\
inverse temperature 
\end{tabular}&
\multicolumn{2}{c}{0.0001}\\
\begin{tabular}{l}
Target value of \\
each inverse temperature 
\end{tabular}&
\multicolumn{2}{c}{5}\\
Number of repetitions & \multicolumn{2}{c}{2}\\
\hline 
\end{tabular}
\end{table}

So far, we have employed the Ising model without the soft constraint.
Next, we investigate the differences in the way of the mapping to the Ising model, i.e., how the formulation with and without constraints makes a difference.
We compare, with respect to decoding accuracy and time, two mappings
(i) the Ising Hamiltonian with soft constraints introducing 
penalty terms corresponding to the syndrome condition Eq.~\eqref{ising_hamiltonian_syndrome_constrain} and (ii) 
the Ising Hamiltonian without the syndrome condition by mapping 
Ising spin variables to stabilizer operators Eq. \eqref{energy function4}.
Here we compare two different mappings in the case of the Pauli $Z$ errors
to make the argument as simple as possible.

Annealing parameters play an important role in optimization performance,
and hence we take the same temperature parameter scheduling for two mappings for a fair comparison
as shown in Tab. \ref{parameter_OpenJij_phaseflip} and \ref{parameter_DA_phaseflip}.
``w" and ``w/o" represent with and without soft constraints, respectively. The columns ``w" of Tab.\ref{parameter_OpenJij_phaseflip} and \ref{parameter_DA_phaseflip} omit $\alpha$ because
there is no need to reduce the degree of the Ising interaction for the method without soft constraints under phase flip noise. 
On the other hand, the lack of Column ``w" of $\alpha$ in Tab. \ref{parameter_DA_phaseflip} stems from employing an alternative method for the degree reduction \cite{fujisaki_2}.

\begin{table}[!t]
\caption{Annealing parameter set of DA}
\label{parameter_DA_phaseflip}
\centering
\begin{tabular}{lcc}
\hline 
Parameter & w & w/o\\
\hline
$J$ & 1 & 1024\\
$h$ & ---  & 1\\
$\alpha$(Equation \ref{penalty_term_depressing}) & --- & ---\\
Annealing mode &  \multicolumn{2}{c}{Replica exchange}\\
Maximum value of temperature & \multicolumn{2}{c}{5}\\
Number of replicas & \multicolumn{2}{c}{128} \\
\hline 
\end{tabular}
\end{table}

\begin{figure}[!t]
    \centering{
        \includegraphics[width=8cm]{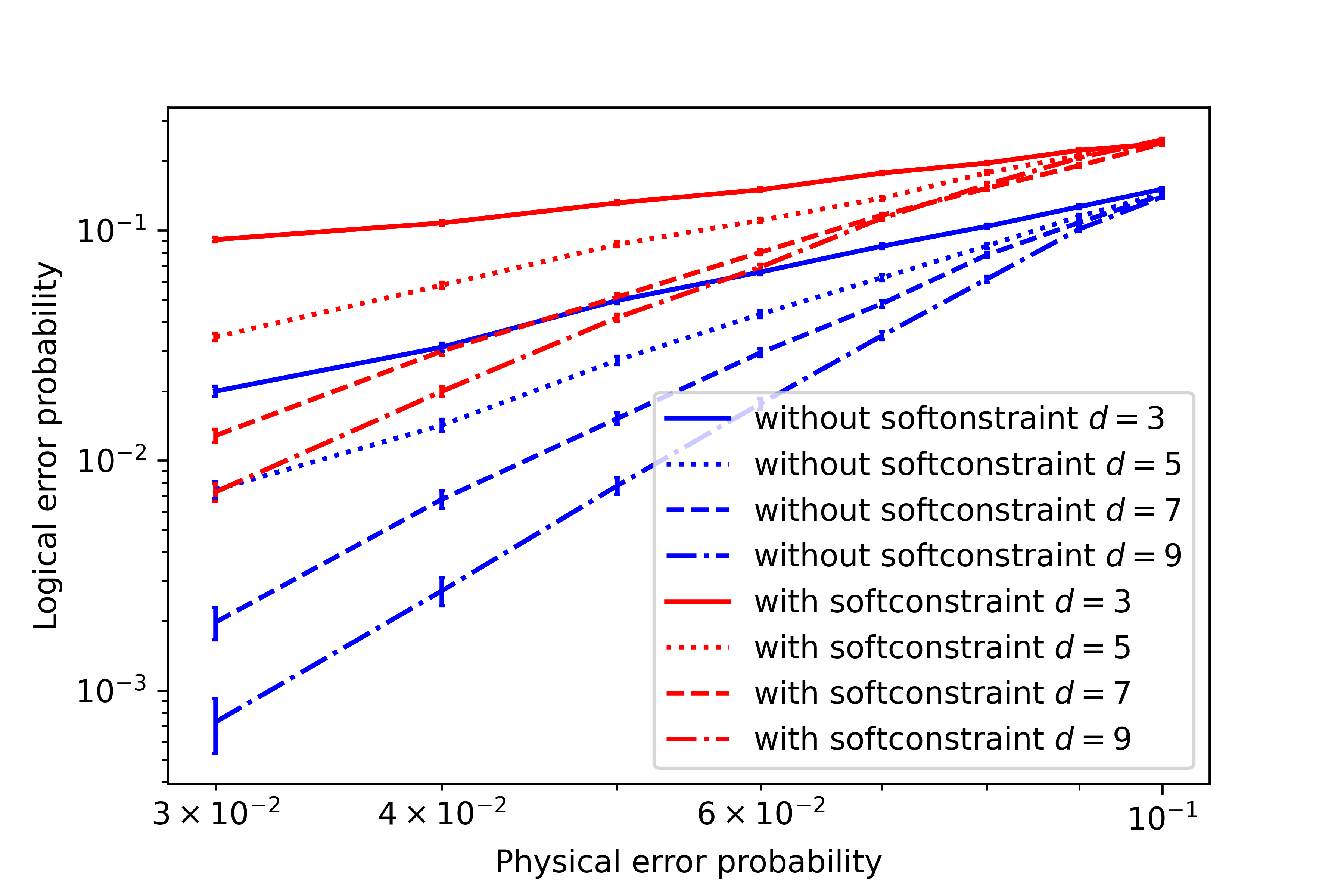}
        \caption{The logical error probability is plotted as a function of the physical error probability with code distances $d=3,5,7,9$ for the SA(OpenJij) decoder with soft constraint(colored red) and the SA(OpenJij) decoder without soft constraints(colored blue).}
        \label{surface_phaseflip_OpenJij}
        }  
\end{figure}

First, we compare two mappings with respect to the logical error probability
using SA(OpenJij).
As shown in Fig. \ref{surface_phaseflip_OpenJij},
mapping (ii) without the soft constraint provides a higher accuracy.
This is expected because the Ising Hamiltonian formulated with the soft constraint is more likely to be trapped in a local minima
due to the complex energy landscape caused by the syndrome constraint.

\begin{figure}[!t]
    \centering{
        \includegraphics[width=8cm]{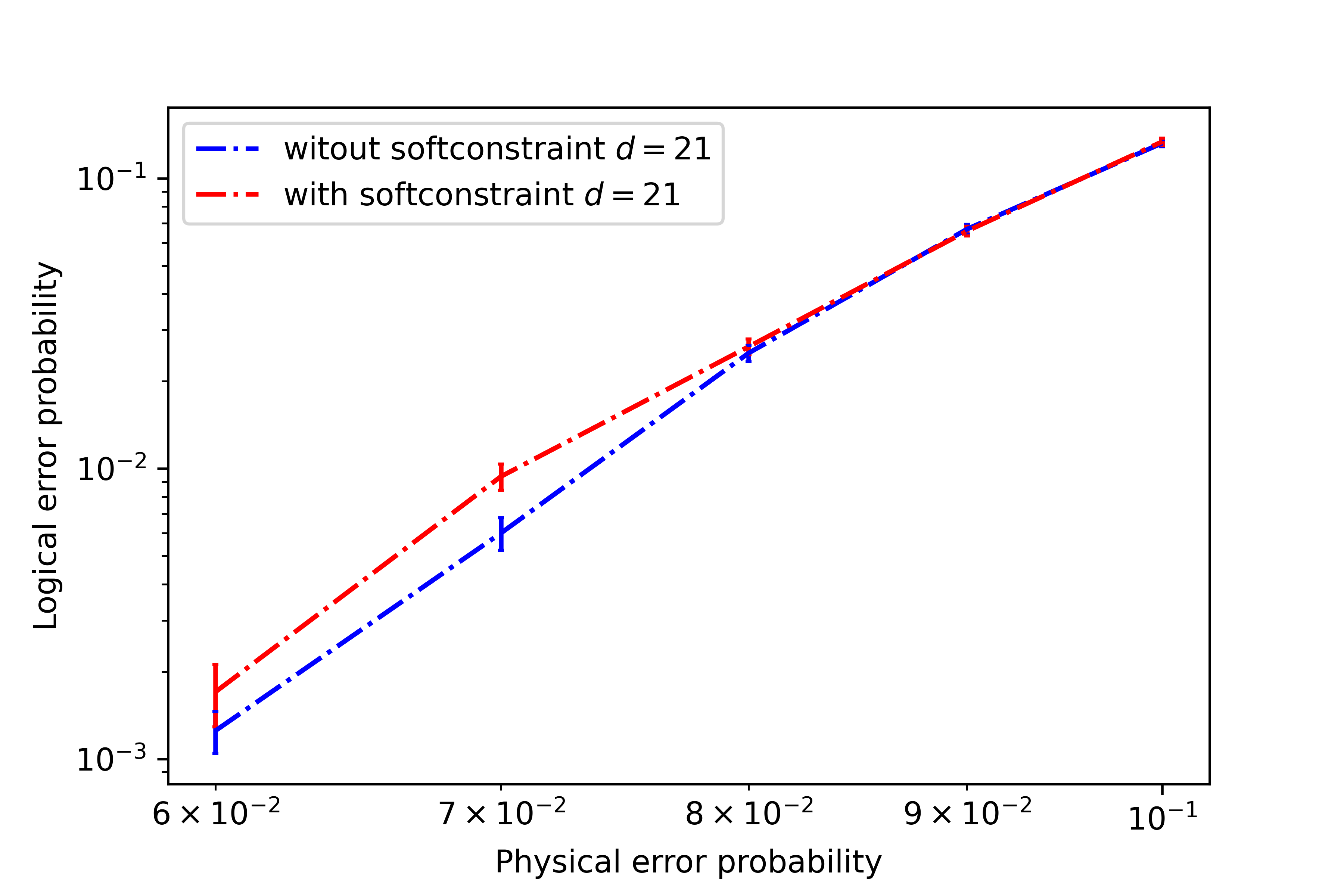}
        \caption{The logical error probability is plotted as a function of the physical error probability with code distance $d=21$ for the DA decoder with soft constraint(colored red) and the DA decoder without soft constraints(colored blue).}
        \label{surface_phaseflip_DA}
        }  
\end{figure}

Next, we perform decoding using DA in the case of $d=21$, using 10000 samples. 
The logical error probability with and without the soft constraint is shown in Fig. \ref{surface_phaseflip_DA}.
While the Ising Hamiltonian without the soft constraint 
provides a slightly smaller logical error probability, two mappings result in almost the same.
This difference in the performance between DA and SA(OpenJij) thought to arise from that 
DA utilizes 
the parallel trial and replica-exchange Monte Carlo for optimization.
At each iteration, DA first searches all possible variables that reduce the cost function or meet Metropolis criterion and chooses one variable that lowers the most energy to update.
Also, unlike SA, the temperature does not monotonically decrease but rises and falls according to a certain rule every few iterations.
It is considered that the optimization avoid falling into a local minimum and make it easier to derive a global minimum solution.
This attributes to the high accuracy even with the soft constraint.


\begin{figure}[b]
    \centering{
        \includegraphics[width=8cm]{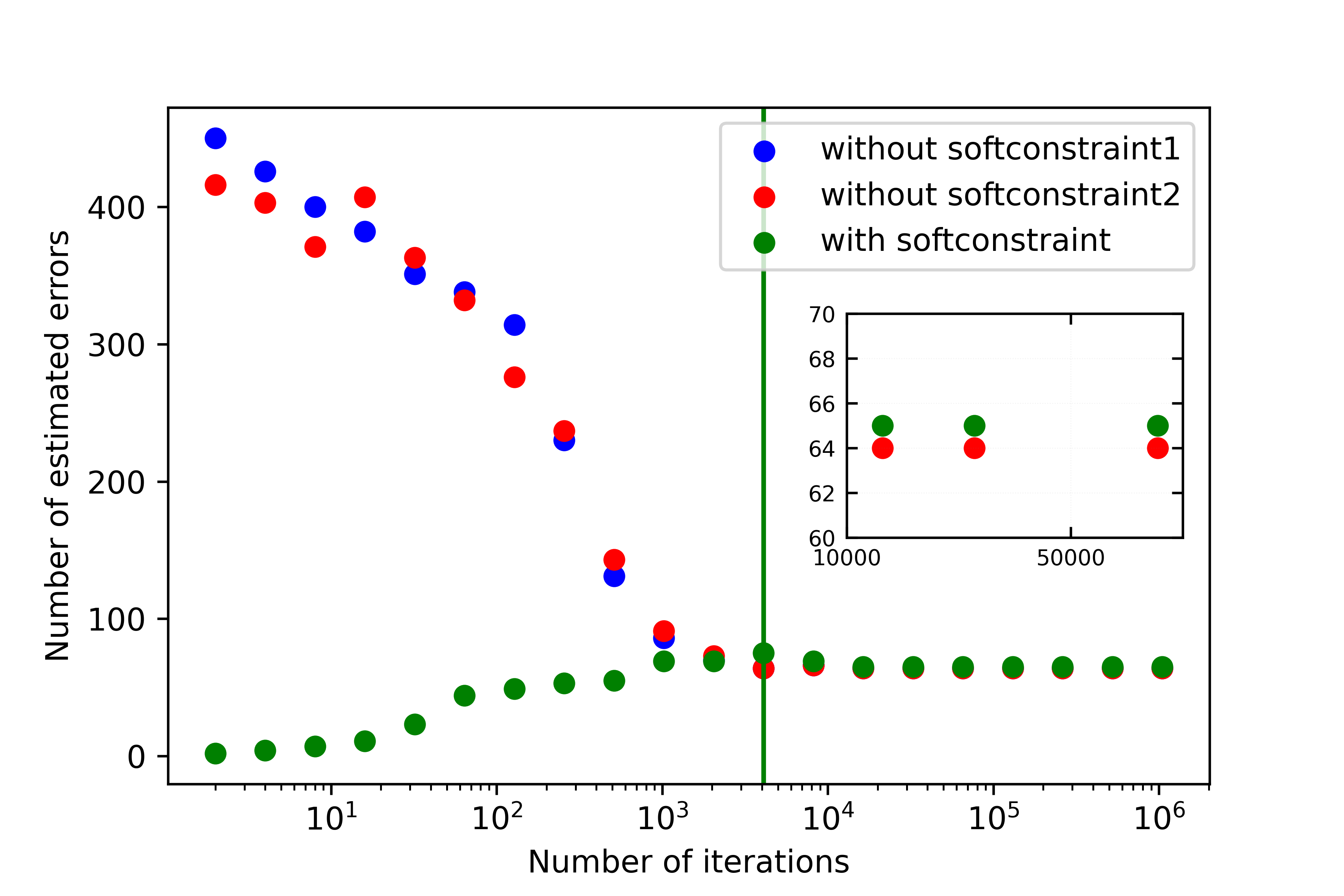}
        \caption{The number of estimated errors is plotted as a function of the number of iterations of DA.
        Without soft constraints, two Hamiltonians (colored by red and blue) with different logical operators are optimized for phase flip noise mode. 
        The green line marks the number of iterations when the soft constraint method first meets the syndrome constraint.}
        \label{num_error0859}
        }  
\end{figure}

To further elucidate the performance differences among two mappings for DA, Fig. \ref{num_error0859} counts 
the number of estimated errors during the annealing process, and Fig. \ref{error_position0859} 
visualizes the estimated error positions. 
Figure \ref{num_error0859} indicates that the Ising Hamiltonian without the soft constraint 
sees a gradual reduction in the error count, allowing for a broader search and preventing logical errors.
On the other hand, 
the error count in the Ising Hamiltonian with the soft constraint remains relatively 
stable after satisfying the soft constraint, i.e., syndrome condition,
meaning trapped at a local minima.
This could lead to a logical error.
We observe the same behavior in various error configurations.

\begin{figure}[!t]
  \begin{minipage}[b]{0.49\linewidth}
    \centering
    \includegraphics[keepaspectratio, scale=0.27]{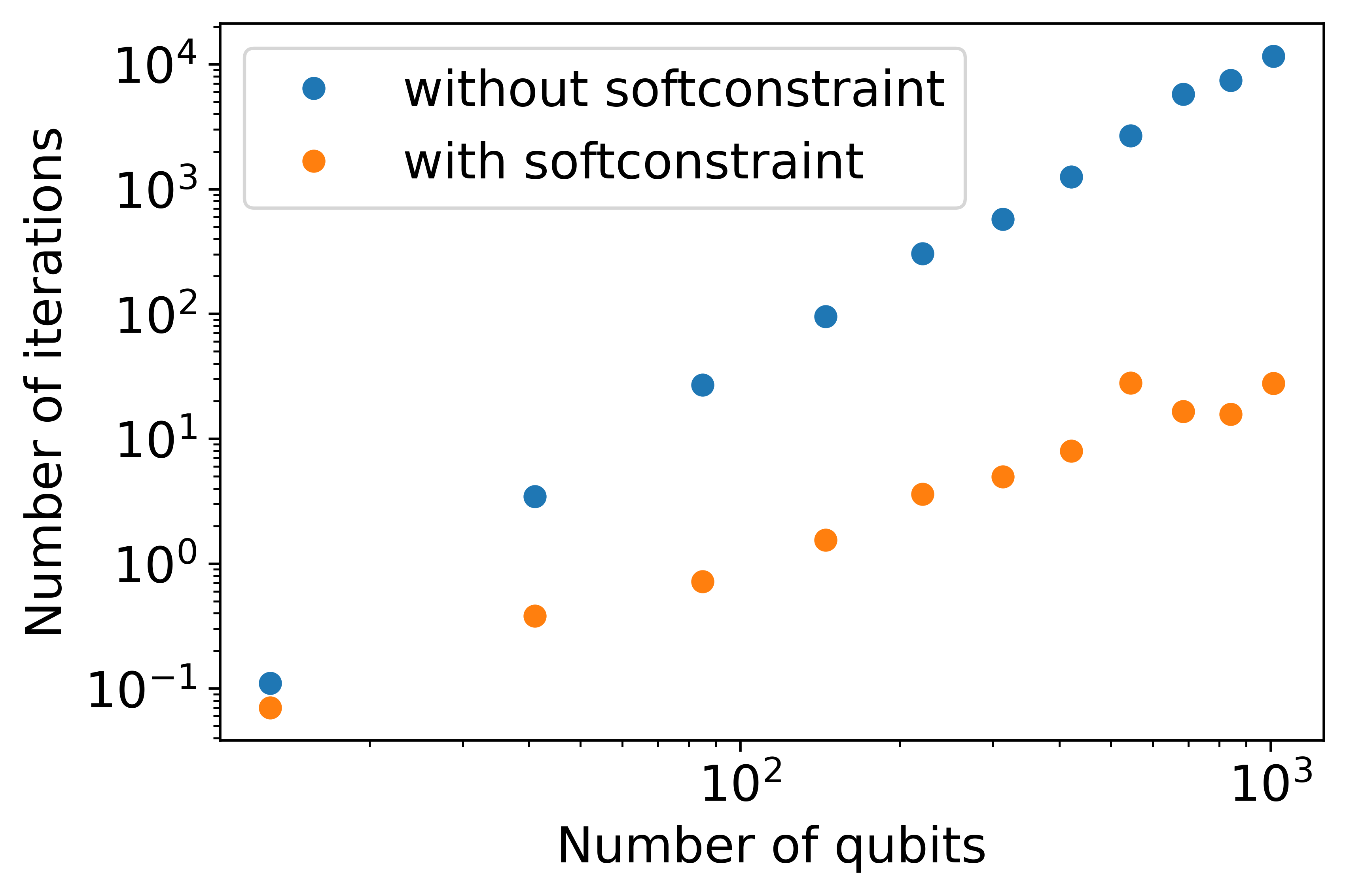}
    \subcaption{}
  \end{minipage}
  \begin{minipage}[b]{0.49\linewidth}
    \centering
    \includegraphics[keepaspectratio, scale=0.27]{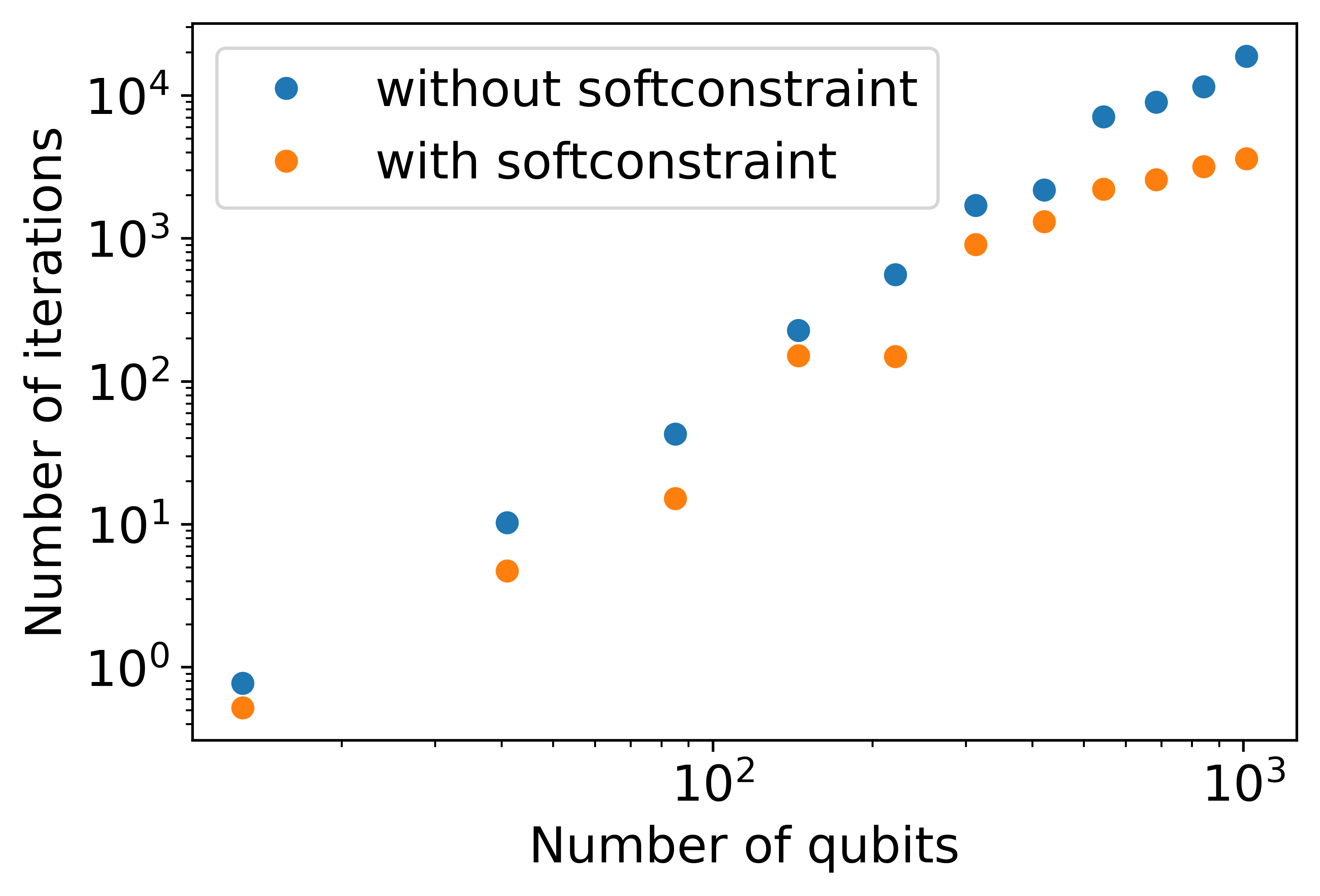}
    \subcaption{}
  \end{minipage}\\
  \begin{minipage}[b]{0.49\linewidth}
    \centering
    \includegraphics[keepaspectratio, scale=0.27]{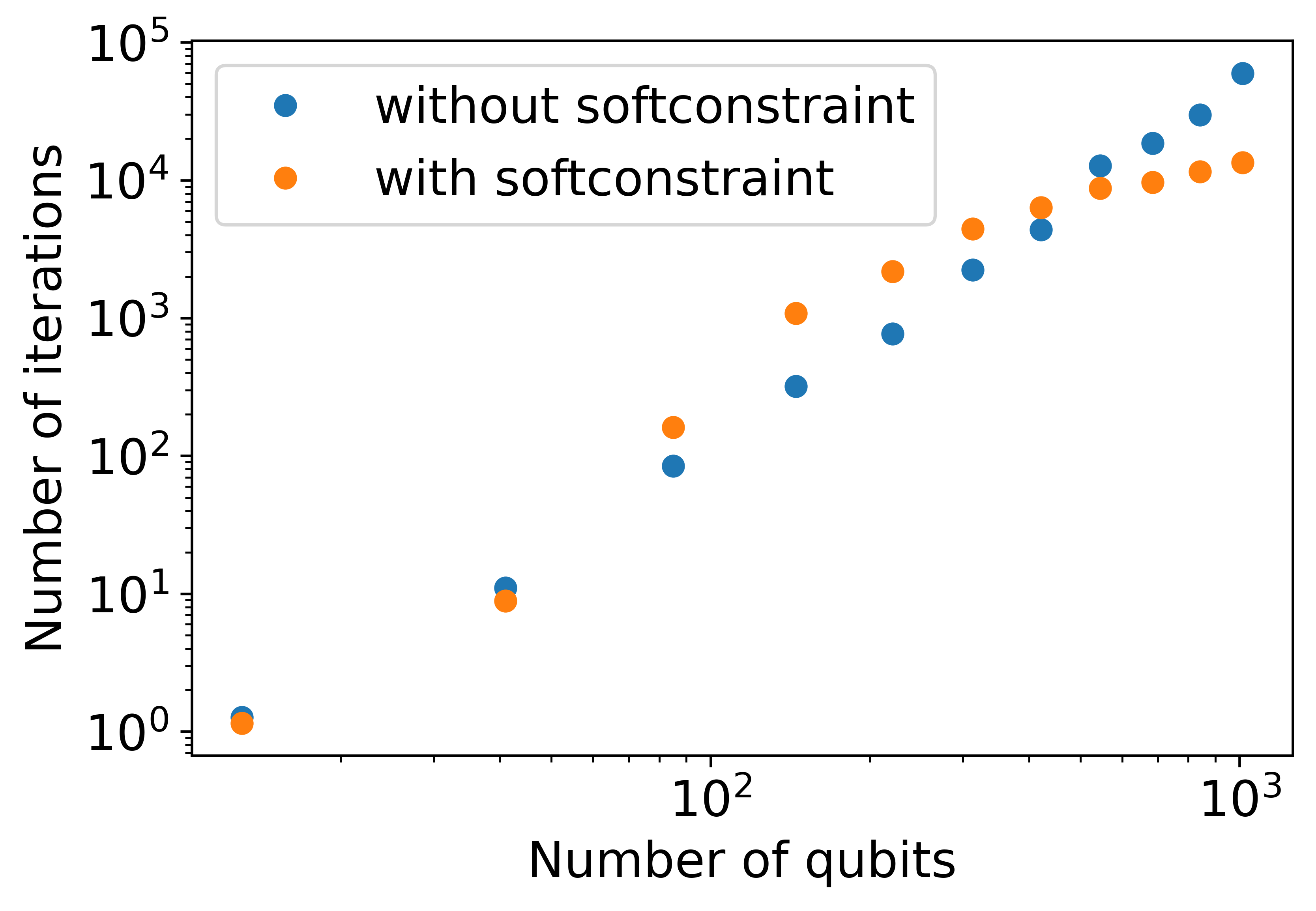}
    \subcaption{}
  \end{minipage}
  \begin{minipage}[b]{0.49\linewidth}
    \centering
    \includegraphics[keepaspectratio, scale=0.27]{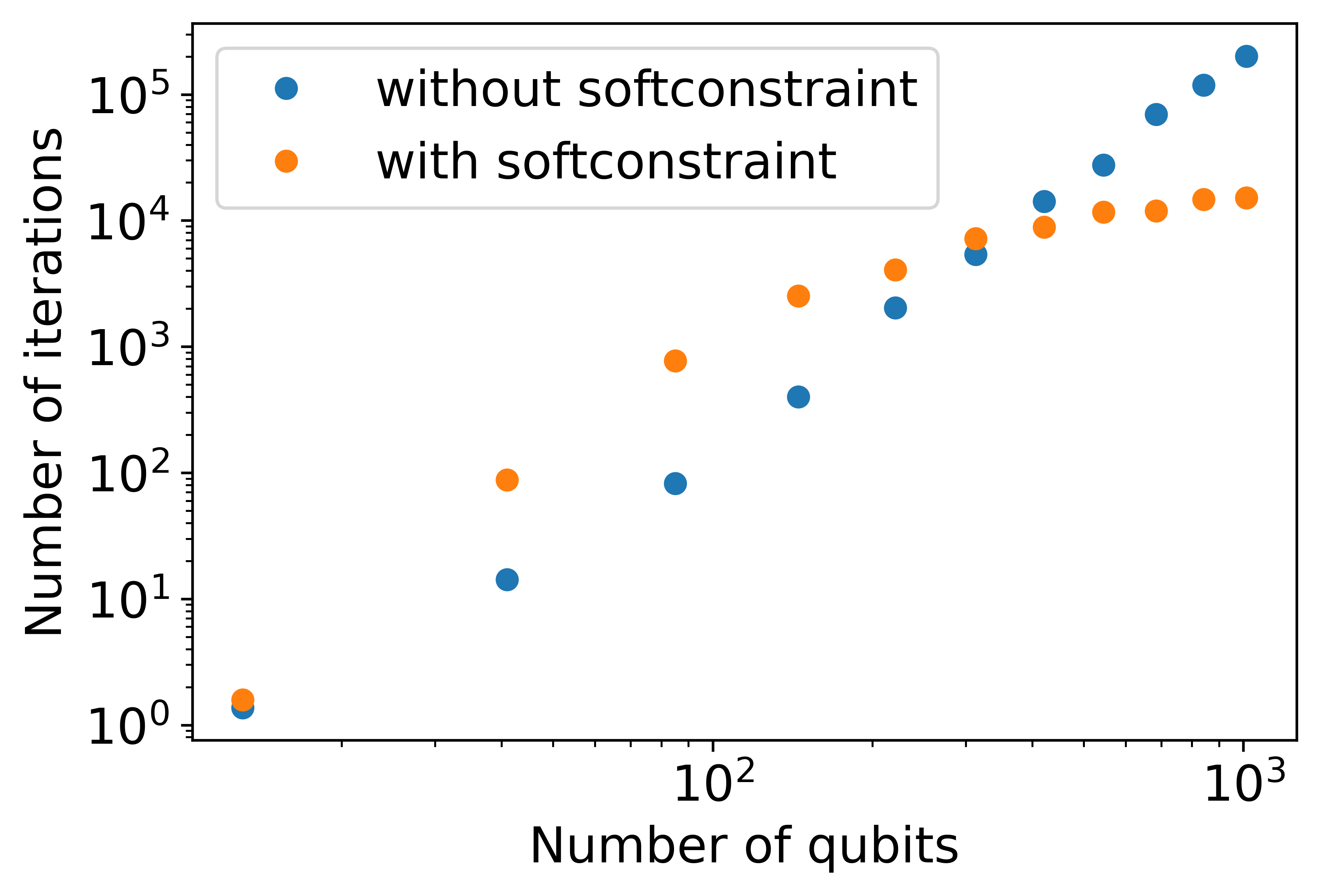}
    \subcaption{}
  \end{minipage}
  \caption{The average number of iterations is plotted as a function of the number of the qubits in the surface codes with physical error probabilities (a) $p=0.01$, (b) $p=0.05$, (c) $p=0.1$ and (d) $p=0.2$.}
  \label{iteration_phaseflip}
\end{figure}

Next, we would like to compare two mappings with respect to the number of iterations to find the solution.
In heuristic algorithms like SA, there is a balance between computation time and solution accuracy. 
More optimization time often yields a solution with lower energy. 
However, for a faster decoding, minimal optimization time achieving reasonably high accuracy is preferred. 
Ideally, optimization process should finish once the lowest energy solution is found. 
We perform decoding for various physical error probabilities and code distances, 
to calculate the average number of iterations required to achieve the lowest energy solution. 
The physical error probabilities ranged from $1{\%}$ to $20{\%}$, and code distances from $3$ to $23$
as in Fig. \ref{iteration_phaseflip}. 
This result indicates that for lower physical error probabilities, methods with the soft constraint consistently require fewer iterations. 
For higher probabilities, especially at larger code distances, the Ising Hamiltonian with soft constraint remains more efficient. 
The following two main factors contribute to these observations:

\begin{figure}[t]
    \centering{
        \includegraphics[width=8cm]{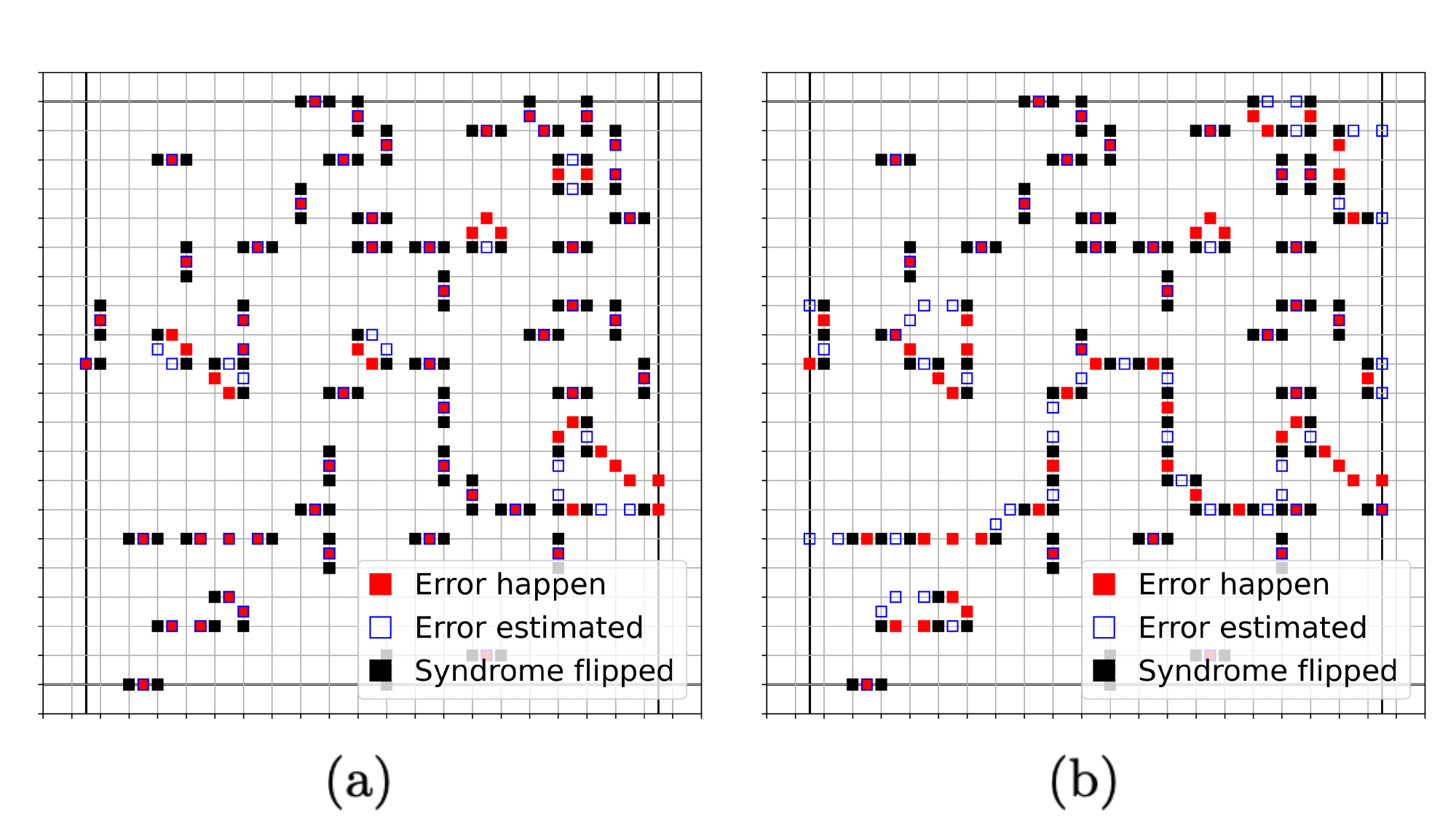}
        \caption{Actual errors (red squares) and estimated errors (blue squares) are visualized for the case with soft constraint (a) and without soft constraint (b). Black squares represent stabilizer operators with their syndrome value $-1$.}
        \label{error_position0859}
        }  
\end{figure}

\begin{itemize}
\item{The impact of soft constraints:} Optimization of the Ising Hamiltonian with the soft constraint tends to 
converge faster while it may be suboptimal with respect to the number of errors 
since priority is given to satisfy the constraint.
In contrast, optimization without the soft constraint continuously search the best solution, with satisfying the syndrome condition, leading to more iterations.

\item{Initial error configurations:} For methods with soft constraints, the initial spin configuration corresponds to no errors. It then searches for configurations with minimal errors while satisfying the syndrome constraint. In methods without soft constraints, while the initial spin configuration already meets the syndrome constraint,
the temperature is chosen to be sufficiently high to search for a globally optimal error configuration. 
\end{itemize}
\noindent It can be seen from the above that in the case of naive SA, 
where no special care is taken against being trapped in a locally optimal solution due to soft constraints, 
it is better to decode using sufficiently long time using embedding without soft constraints. 
Even though, as for scaling, since IP(CPLEX) increases exponentially, SA will eventually achieve higher performance with comparable accuracy as the code distance increases. 
On the other hand, more sophisticated approach amenable to the soft constrain, such as DA using replica exchange Monte Calro, 
it is much faster to find a solution with fewer iterations in the presence of the soft constraint. 
In particular, when the error probability is low and the code distance is large, 
the DA decoder with a formulation with the soft constraint 
performs best. Since it takes about $1\mu$sec per iteration~\cite{da_annealing_time}, 
the decoding time is in tens of milliseconds, even when the number of qubits is $\sim 10^3$. 
This is significantly faster than IP (CPLEX) and SA (OpenJij), which can achieve the same high decoding accuracy, and even comparable to the MWPM decoder with the code distance hitting the same logical error rates.

\section{Conclusion and Discussion}
\label{Conclusion_and_Discussion}
This study investigated the surface code decoding under depolarizing noise 
using Ising model solvers. 
We found that the SA and DA decoders exhibit higher decoding accuracy under depolarizing noise compared to MWPM. 
Additionally, near threshold physical error probabilities, 
our method decodes faster than the IP decoder using CPLEX, 
maintaining comparable accuracy. 
While the SA decoder using OpenJij cannot surpass the MWPM decoder,
the Ising-based decoders are simple and hence 
has high flexibility for their implementation on a massively parallel processor or 
dedicated hardware close to a quantum device.
For example, we can implement the SA decoder on FPGA with 
smaller latency for the decoding,
since FPGA-based Ising solver implementations have already been reported (\cite{FPGA1},\cite{FPGA2}). 
We have also explored the formulations of the Ising models for decoding,
with and without the soft constraint.
Our result has indicated the trade-off between optimization accuracy and computational complexity, emphasizing the need for application-specific choices. 
Also depending on the Ising solver, the impact of with and without the soft constraint is quite different.

While our evaluations assumed error-free syndrome measurements,
a realistic quantum computing scenario involves 
syndrome measurement errors. 
As done in Ref.~\cite{takada}, an extension of the proposed method for the 
imperfect syndrome measurement is straightforward.
The versatility of our approach, applicable to various stabilizer codes, suggests potential broader applications, such as quantum LDPC codes~\cite{LDPC_IBM,LDPC_Lukin}.

\begin{acknowledgments}
This work is supported by MEXT Quantum Leap Flagship Program (MEXT Q-LEAP) Grant No. JPMXS0118067394 and JPMXS0120319794, 
JST COI-NEXT Grant No. JPMJPF2014, and JST Moonshot R\&D Grant No. JPMJMS2061.
\end{acknowledgments}

\bibliography{main}

\begin{thebibliography}{31}%
\makeatletter
\providecommand \@ifxundefined [1]{%
 \@ifx{#1\undefined}
}%
\providecommand \@ifnum [1]{%
 \ifnum #1\expandafter \@firstoftwo
 \else \expandafter \@secondoftwo
 \fi
}%
\providecommand \@ifx [1]{%
 \ifx #1\expandafter \@firstoftwo
 \else \expandafter \@secondoftwo
 \fi
}%
\providecommand \natexlab [1]{#1}%
\providecommand \enquote  [1]{``#1''}%
\providecommand \bibnamefont  [1]{#1}%
\providecommand \bibfnamefont [1]{#1}%
\providecommand \citenamefont [1]{#1}%
\providecommand \href@noop [0]{\@secondoftwo}%
\providecommand \href [0]{\begingroup \@sanitize@url \@href}%
\providecommand \@href[1]{\@@startlink{#1}\@@href}%
\providecommand \@@href[1]{\endgroup#1\@@endlink}%
\providecommand \@sanitize@url [0]{\catcode `\\12\catcode `\$12\catcode `\&12\catcode `\#12\catcode `\^12\catcode `\_12\catcode `\%12\relax}%
\providecommand \@@startlink[1]{}%
\providecommand \@@endlink[0]{}%
\providecommand \url  [0]{\begingroup\@sanitize@url \@url }%
\providecommand \@url [1]{\endgroup\@href {#1}{\urlprefix }}%
\providecommand \urlprefix  [0]{URL }%
\providecommand \Eprint [0]{\href }%
\providecommand \doibase [0]{http://dx.doi.org/}%
\providecommand \selectlanguage [0]{\@gobble}%
\providecommand \bibinfo  [0]{\@secondoftwo}%
\providecommand \bibfield  [0]{\@secondoftwo}%
\providecommand \translation [1]{[#1]}%
\providecommand \BibitemOpen [0]{}%
\providecommand \bibitemStop [0]{}%
\providecommand \bibitemNoStop [0]{.\EOS\space}%
\providecommand \EOS [0]{\spacefactor3000\relax}%
\providecommand \BibitemShut  [1]{\csname bibitem#1\endcsname}%
\let\auto@bib@innerbib\@empty
\bibitem [{\citenamefont {Fujii}(2015)}]{fujii_topological_code}%
  \BibitemOpen
  \bibfield  {author} {\bibinfo {author} {\bibfnamefont {Keisuke}\ \bibnamefont {Fujii}},\ }\href@noop {} {\emph {\bibinfo {title} {Quantum Computation with Topological Codes: from qubit to topological fault-tolerance}}},\ Vol.~\bibinfo {volume} {8}\ (\bibinfo  {publisher} {Springer},\ \bibinfo {year} {2015})\BibitemShut {NoStop}%
\bibitem [{goo(2023)}]{google_supremacy}%
  \BibitemOpen
  \bibfield  {title} {\enquote {\bibinfo {title} {Suppressing quantum errors by scaling a surface code logical qubit},}\ }\href@noop {} {\bibfield  {journal} {\bibinfo  {journal} {Nature}\ }\textbf {\bibinfo {volume} {614}},\ \bibinfo {pages} {676--681} (\bibinfo {year} {2023})}\BibitemShut {NoStop}%
\bibitem [{\citenamefont {Kim}\ \emph {et~al.}(2023)\citenamefont {Kim}, \citenamefont {Eddins}, \citenamefont {Anand}, \citenamefont {Wei}, \citenamefont {van~den Berg}, \citenamefont {Rosenblatt}, \citenamefont {Nayfeh}, \citenamefont {Wu}, \citenamefont {Zaletel}, \citenamefont {Temme},\ and\ \citenamefont {Kandala}}]{IBM_utility}%
  \BibitemOpen
  \bibfield  {author} {\bibinfo {author} {\bibfnamefont {Youngseok}\ \bibnamefont {Kim}}, \bibinfo {author} {\bibfnamefont {Andrew}\ \bibnamefont {Eddins}}, \bibinfo {author} {\bibfnamefont {Sajant}\ \bibnamefont {Anand}}, \bibinfo {author} {\bibfnamefont {Ken~Xuan}\ \bibnamefont {Wei}}, \bibinfo {author} {\bibfnamefont {Ewout}\ \bibnamefont {van~den Berg}}, \bibinfo {author} {\bibfnamefont {Sami}\ \bibnamefont {Rosenblatt}}, \bibinfo {author} {\bibfnamefont {Hasan}\ \bibnamefont {Nayfeh}}, \bibinfo {author} {\bibfnamefont {Yantao}\ \bibnamefont {Wu}}, \bibinfo {author} {\bibfnamefont {Michael}\ \bibnamefont {Zaletel}}, \bibinfo {author} {\bibfnamefont {Kristan}\ \bibnamefont {Temme}}, \ and\ \bibinfo {author} {\bibfnamefont {Abhinav}\ \bibnamefont {Kandala}},\ }\bibfield  {title} {\enquote {\bibinfo {title} {Evidence for the utility of quantum computing before fault tolerance},}\ }\href {\doibase 10.1038/s41586-023-06096-3} {\bibfield  {journal} {\bibinfo  {journal} {Nature}\ }\textbf {\bibinfo {volume}
  {618}},\ \bibinfo {pages} {500--505} (\bibinfo {year} {2023})}\BibitemShut {NoStop}%
\bibitem [{\citenamefont {Chen}\ \emph {et~al.}(2021)\citenamefont {Chen}, \citenamefont {Satzinger}, \citenamefont {Atalaya}, \citenamefont {Korotkov}, \citenamefont {Dunsworth}, \citenamefont {Sank}, \citenamefont {Quintana}, \citenamefont {McEwen}, \citenamefont {Barends}, \citenamefont {Klimov}, \citenamefont {Hong}, \citenamefont {Jones}, \citenamefont {Petukhov}, \citenamefont {Kafri}, \citenamefont {Demura}, \citenamefont {Burkett}, \citenamefont {Gidney}, \citenamefont {Fowler}, \citenamefont {Paler}, \citenamefont {Putterman}, \citenamefont {Aleiner}, \citenamefont {Arute}, \citenamefont {Arya}, \citenamefont {Babbush}, \citenamefont {Bardin}, \citenamefont {Bengtsson}, \citenamefont {Bourassa}, \citenamefont {Broughton}, \citenamefont {Buckley}, \citenamefont {Buell}, \citenamefont {Bushnell}, \citenamefont {Chiaro}, \citenamefont {Collins}, \citenamefont {Courtney}, \citenamefont {Derk}, \citenamefont {Eppens}, \citenamefont {Erickson}, \citenamefont {Farhi}, \citenamefont {Foxen}, \citenamefont
  {Giustina}, \citenamefont {Greene}, \citenamefont {Gross}, \citenamefont {Harrigan}, \citenamefont {Harrington}, \citenamefont {Hilton}, \citenamefont {Ho}, \citenamefont {Huang}, \citenamefont {Huggins}, \citenamefont {Ioffe}, \citenamefont {Isakov}, \citenamefont {Jeffrey}, \citenamefont {Jiang}, \citenamefont {Kechedzhi}, \citenamefont {Kim}, \citenamefont {Kitaev}, \citenamefont {Kostritsa}, \citenamefont {Landhuis}, \citenamefont {Laptev}, \citenamefont {Lucero}, \citenamefont {Martin}, \citenamefont {McClean}, \citenamefont {McCourt}, \citenamefont {Mi}, \citenamefont {Miao}, \citenamefont {Mohseni}, \citenamefont {Montazeri}, \citenamefont {Mruczkiewicz}, \citenamefont {Mutus}, \citenamefont {Naaman}, \citenamefont {Neeley}, \citenamefont {Neill}, \citenamefont {Newman}, \citenamefont {Niu}, \citenamefont {O'Brien}, \citenamefont {Opremcak}, \citenamefont {Ostby}, \citenamefont {Pat{\'o}}, \citenamefont {Redd}, \citenamefont {Roushan}, \citenamefont {Rubin}, \citenamefont {Shvarts}, \citenamefont
  {Strain}, \citenamefont {Szalay}, \citenamefont {Trevithick}, \citenamefont {Villalonga}, \citenamefont {White}, \citenamefont {Yao}, \citenamefont {Yeh}, \citenamefont {Yoo}, \citenamefont {Zalcman}, \citenamefont {Neven}, \citenamefont {Boixo}, \citenamefont {Smelyanskiy}, \citenamefont {Chen}, \citenamefont {Megrant}, \citenamefont {Kelly},\ and\ \citenamefont {AI}}]{google_qec}%
  \BibitemOpen
  \bibfield  {author} {\bibinfo {author} {\bibfnamefont {Zijun}\ \bibnamefont {Chen}}, \bibinfo {author} {\bibfnamefont {Kevin~J.}\ \bibnamefont {Satzinger}}, \bibinfo {author} {\bibfnamefont {Juan}\ \bibnamefont {Atalaya}}, \bibinfo {author} {\bibfnamefont {Alexander~N.}\ \bibnamefont {Korotkov}}, \bibinfo {author} {\bibfnamefont {Andrew}\ \bibnamefont {Dunsworth}}, \bibinfo {author} {\bibfnamefont {Daniel}\ \bibnamefont {Sank}}, \bibinfo {author} {\bibfnamefont {Chris}\ \bibnamefont {Quintana}}, \bibinfo {author} {\bibfnamefont {Matt}\ \bibnamefont {McEwen}}, \bibinfo {author} {\bibfnamefont {Rami}\ \bibnamefont {Barends}}, \bibinfo {author} {\bibfnamefont {Paul~V.}\ \bibnamefont {Klimov}}, \bibinfo {author} {\bibfnamefont {Sabrina}\ \bibnamefont {Hong}}, \bibinfo {author} {\bibfnamefont {Cody}\ \bibnamefont {Jones}}, \bibinfo {author} {\bibfnamefont {Andre}\ \bibnamefont {Petukhov}}, \bibinfo {author} {\bibfnamefont {Dvir}\ \bibnamefont {Kafri}}, \bibinfo {author} {\bibfnamefont {Sean}\ \bibnamefont
  {Demura}}, \bibinfo {author} {\bibfnamefont {Brian}\ \bibnamefont {Burkett}}, \bibinfo {author} {\bibfnamefont {Craig}\ \bibnamefont {Gidney}}, \bibinfo {author} {\bibfnamefont {Austin~G.}\ \bibnamefont {Fowler}}, \bibinfo {author} {\bibfnamefont {Alexandru}\ \bibnamefont {Paler}}, \bibinfo {author} {\bibfnamefont {Harald}\ \bibnamefont {Putterman}}, \bibinfo {author} {\bibfnamefont {Igor}\ \bibnamefont {Aleiner}}, \bibinfo {author} {\bibfnamefont {Frank}\ \bibnamefont {Arute}}, \bibinfo {author} {\bibfnamefont {Kunal}\ \bibnamefont {Arya}}, \bibinfo {author} {\bibfnamefont {Ryan}\ \bibnamefont {Babbush}}, \bibinfo {author} {\bibfnamefont {Joseph~C.}\ \bibnamefont {Bardin}}, \bibinfo {author} {\bibfnamefont {Andreas}\ \bibnamefont {Bengtsson}}, \bibinfo {author} {\bibfnamefont {Alexandre}\ \bibnamefont {Bourassa}}, \bibinfo {author} {\bibfnamefont {Michael}\ \bibnamefont {Broughton}}, \bibinfo {author} {\bibfnamefont {Bob~B.}\ \bibnamefont {Buckley}}, \bibinfo {author} {\bibfnamefont {David~A.}\
  \bibnamefont {Buell}}, \bibinfo {author} {\bibfnamefont {Nicholas}\ \bibnamefont {Bushnell}}, \bibinfo {author} {\bibfnamefont {Benjamin}\ \bibnamefont {Chiaro}}, \bibinfo {author} {\bibfnamefont {Roberto}\ \bibnamefont {Collins}}, \bibinfo {author} {\bibfnamefont {William}\ \bibnamefont {Courtney}}, \bibinfo {author} {\bibfnamefont {Alan~R.}\ \bibnamefont {Derk}}, \bibinfo {author} {\bibfnamefont {Daniel}\ \bibnamefont {Eppens}}, \bibinfo {author} {\bibfnamefont {Catherine}\ \bibnamefont {Erickson}}, \bibinfo {author} {\bibfnamefont {Edward}\ \bibnamefont {Farhi}}, \bibinfo {author} {\bibfnamefont {Brooks}\ \bibnamefont {Foxen}}, \bibinfo {author} {\bibfnamefont {Marissa}\ \bibnamefont {Giustina}}, \bibinfo {author} {\bibfnamefont {Ami}\ \bibnamefont {Greene}}, \bibinfo {author} {\bibfnamefont {Jonathan~A.}\ \bibnamefont {Gross}}, \bibinfo {author} {\bibfnamefont {Matthew~P.}\ \bibnamefont {Harrigan}}, \bibinfo {author} {\bibfnamefont {Sean~D.}\ \bibnamefont {Harrington}}, \bibinfo {author} {\bibfnamefont
  {Jeremy}\ \bibnamefont {Hilton}}, \bibinfo {author} {\bibfnamefont {Alan}\ \bibnamefont {Ho}}, \bibinfo {author} {\bibfnamefont {Trent}\ \bibnamefont {Huang}}, \bibinfo {author} {\bibfnamefont {William~J.}\ \bibnamefont {Huggins}}, \bibinfo {author} {\bibfnamefont {L.~B.}\ \bibnamefont {Ioffe}}, \bibinfo {author} {\bibfnamefont {Sergei~V.}\ \bibnamefont {Isakov}}, \bibinfo {author} {\bibfnamefont {Evan}\ \bibnamefont {Jeffrey}}, \bibinfo {author} {\bibfnamefont {Zhang}\ \bibnamefont {Jiang}}, \bibinfo {author} {\bibfnamefont {Kostyantyn}\ \bibnamefont {Kechedzhi}}, \bibinfo {author} {\bibfnamefont {Seon}\ \bibnamefont {Kim}}, \bibinfo {author} {\bibfnamefont {Alexei}\ \bibnamefont {Kitaev}}, \bibinfo {author} {\bibfnamefont {Fedor}\ \bibnamefont {Kostritsa}}, \bibinfo {author} {\bibfnamefont {David}\ \bibnamefont {Landhuis}}, \bibinfo {author} {\bibfnamefont {Pavel}\ \bibnamefont {Laptev}}, \bibinfo {author} {\bibfnamefont {Erik}\ \bibnamefont {Lucero}}, \bibinfo {author} {\bibfnamefont {Orion}\
  \bibnamefont {Martin}}, \bibinfo {author} {\bibfnamefont {Jarrod~R.}\ \bibnamefont {McClean}}, \bibinfo {author} {\bibfnamefont {Trevor}\ \bibnamefont {McCourt}}, \bibinfo {author} {\bibfnamefont {Xiao}\ \bibnamefont {Mi}}, \bibinfo {author} {\bibfnamefont {Kevin~C.}\ \bibnamefont {Miao}}, \bibinfo {author} {\bibfnamefont {Masoud}\ \bibnamefont {Mohseni}}, \bibinfo {author} {\bibfnamefont {Shirin}\ \bibnamefont {Montazeri}}, \bibinfo {author} {\bibfnamefont {Wojciech}\ \bibnamefont {Mruczkiewicz}}, \bibinfo {author} {\bibfnamefont {Josh}\ \bibnamefont {Mutus}}, \bibinfo {author} {\bibfnamefont {Ofer}\ \bibnamefont {Naaman}}, \bibinfo {author} {\bibfnamefont {Matthew}\ \bibnamefont {Neeley}}, \bibinfo {author} {\bibfnamefont {Charles}\ \bibnamefont {Neill}}, \bibinfo {author} {\bibfnamefont {Michael}\ \bibnamefont {Newman}}, \bibinfo {author} {\bibfnamefont {Murphy~Yuezhen}\ \bibnamefont {Niu}}, \bibinfo {author} {\bibfnamefont {Thomas~E.}\ \bibnamefont {O'Brien}}, \bibinfo {author} {\bibfnamefont {Alex}\
  \bibnamefont {Opremcak}}, \bibinfo {author} {\bibfnamefont {Eric}\ \bibnamefont {Ostby}}, \bibinfo {author} {\bibfnamefont {B{\'a}lint}\ \bibnamefont {Pat{\'o}}}, \bibinfo {author} {\bibfnamefont {Nicholas}\ \bibnamefont {Redd}}, \bibinfo {author} {\bibfnamefont {Pedram}\ \bibnamefont {Roushan}}, \bibinfo {author} {\bibfnamefont {Nicholas~C.}\ \bibnamefont {Rubin}}, \bibinfo {author} {\bibfnamefont {Vladimir}\ \bibnamefont {Shvarts}}, \bibinfo {author} {\bibfnamefont {Doug}\ \bibnamefont {Strain}}, \bibinfo {author} {\bibfnamefont {Marco}\ \bibnamefont {Szalay}}, \bibinfo {author} {\bibfnamefont {Matthew~D.}\ \bibnamefont {Trevithick}}, \bibinfo {author} {\bibfnamefont {Benjamin}\ \bibnamefont {Villalonga}}, \bibinfo {author} {\bibfnamefont {Theodore}\ \bibnamefont {White}}, \bibinfo {author} {\bibfnamefont {Z.~Jamie}\ \bibnamefont {Yao}}, \bibinfo {author} {\bibfnamefont {Ping}\ \bibnamefont {Yeh}}, \bibinfo {author} {\bibfnamefont {Juhwan}\ \bibnamefont {Yoo}}, \bibinfo {author} {\bibfnamefont {Adam}\
  \bibnamefont {Zalcman}}, \bibinfo {author} {\bibfnamefont {Hartmut}\ \bibnamefont {Neven}}, \bibinfo {author} {\bibfnamefont {Sergio}\ \bibnamefont {Boixo}}, \bibinfo {author} {\bibfnamefont {Vadim}\ \bibnamefont {Smelyanskiy}}, \bibinfo {author} {\bibfnamefont {Yu}~\bibnamefont {Chen}}, \bibinfo {author} {\bibfnamefont {Anthony}\ \bibnamefont {Megrant}}, \bibinfo {author} {\bibfnamefont {Julian}\ \bibnamefont {Kelly}}, \ and\ \bibinfo {author} {\bibfnamefont {Google~Quantum}\ \bibnamefont {AI}},\ }\bibfield  {title} {\enquote {\bibinfo {title} {Exponential suppression of bit or phase errors with cyclic error correction},}\ }\href {\doibase 10.1038/s41586-021-03588-y} {\bibfield  {journal} {\bibinfo  {journal} {Nature}\ }\textbf {\bibinfo {volume} {595}},\ \bibinfo {pages} {383--387} (\bibinfo {year} {2021})}\BibitemShut {NoStop}%
\bibitem [{\citenamefont {Gidney}\ and\ \citenamefont {Eker{\aa}}(2021)}]{2048_factoring}%
  \BibitemOpen
  \bibfield  {author} {\bibinfo {author} {\bibfnamefont {Craig}\ \bibnamefont {Gidney}}\ and\ \bibinfo {author} {\bibfnamefont {Martin}\ \bibnamefont {Eker{\aa}}},\ }\bibfield  {title} {\enquote {\bibinfo {title} {How to factor 2048 bit rsa integers in 8 hours using 20 million noisy qubits},}\ }\href@noop {} {\bibfield  {journal} {\bibinfo  {journal} {Quantum}\ }\textbf {\bibinfo {volume} {5}},\ \bibinfo {pages} {433} (\bibinfo {year} {2021})}\BibitemShut {NoStop}%
\bibitem [{\citenamefont {Yoshioka}\ \emph {et~al.}(2022)\citenamefont {Yoshioka}, \citenamefont {Okubo}, \citenamefont {Suzuki}, \citenamefont {Koizumi},\ and\ \citenamefont {Mizukami}}]{Yoshioka_Mizukami}%
  \BibitemOpen
  \bibfield  {author} {\bibinfo {author} {\bibfnamefont {Nobuyuki}\ \bibnamefont {Yoshioka}}, \bibinfo {author} {\bibfnamefont {Tsuyoshi}\ \bibnamefont {Okubo}}, \bibinfo {author} {\bibfnamefont {Yasunari}\ \bibnamefont {Suzuki}}, \bibinfo {author} {\bibfnamefont {Yuki}\ \bibnamefont {Koizumi}}, \ and\ \bibinfo {author} {\bibfnamefont {Wataru}\ \bibnamefont {Mizukami}},\ }\bibfield  {title} {\enquote {\bibinfo {title} {Hunting for quantum-classical crossover in condensed matter problems},}\ }\href@noop {} {\bibfield  {journal} {\bibinfo  {journal} {arXiv preprint arXiv:2210.14109}\ } (\bibinfo {year} {2022})}\BibitemShut {NoStop}%
\bibitem [{\citenamefont {Edmonds}(1965)}]{MWPM}%
  \BibitemOpen
  \bibfield  {author} {\bibinfo {author} {\bibfnamefont {Jack}\ \bibnamefont {Edmonds}},\ }\bibfield  {title} {\enquote {\bibinfo {title} {Paths, trees, and flowers},}\ }\href@noop {} {\bibfield  {journal} {\bibinfo  {journal} {Canadian Journal of mathematics}\ }\textbf {\bibinfo {volume} {17}},\ \bibinfo {pages} {449--467} (\bibinfo {year} {1965})}\BibitemShut {NoStop}%
\bibitem [{\citenamefont {Fowler}\ \emph {et~al.}(2012)\citenamefont {Fowler}, \citenamefont {Mariantoni}, \citenamefont {Martinis},\ and\ \citenamefont {Cleland}}]{surface_Fowler}%
  \BibitemOpen
  \bibfield  {author} {\bibinfo {author} {\bibfnamefont {Austin~G}\ \bibnamefont {Fowler}}, \bibinfo {author} {\bibfnamefont {Matteo}\ \bibnamefont {Mariantoni}}, \bibinfo {author} {\bibfnamefont {John~M}\ \bibnamefont {Martinis}}, \ and\ \bibinfo {author} {\bibfnamefont {Andrew~N}\ \bibnamefont {Cleland}},\ }\bibfield  {title} {\enquote {\bibinfo {title} {Surface codes: Towards practical large-scale quantum computation},}\ }\href@noop {} {\bibfield  {journal} {\bibinfo  {journal} {Physical Review A}\ }\textbf {\bibinfo {volume} {86}},\ \bibinfo {pages} {032324} (\bibinfo {year} {2012})}\BibitemShut {NoStop}%
\bibitem [{\citenamefont {Davaasuren}\ \emph {et~al.}(2020)\citenamefont {Davaasuren}, \citenamefont {Suzuki}, \citenamefont {Fujii},\ and\ \citenamefont {Koashi}}]{Suzuki}%
  \BibitemOpen
  \bibfield  {author} {\bibinfo {author} {\bibfnamefont {Amarsanaa}\ \bibnamefont {Davaasuren}}, \bibinfo {author} {\bibfnamefont {Yasunari}\ \bibnamefont {Suzuki}}, \bibinfo {author} {\bibfnamefont {Keisuke}\ \bibnamefont {Fujii}}, \ and\ \bibinfo {author} {\bibfnamefont {Masato}\ \bibnamefont {Koashi}},\ }\bibfield  {title} {\enquote {\bibinfo {title} {General framework for constructing fast and near-optimal machine-learning-based decoder of the topological stabilizer codes},}\ }\href@noop {} {\bibfield  {journal} {\bibinfo  {journal} {Physical Review Research}\ }\textbf {\bibinfo {volume} {2}},\ \bibinfo {pages} {033399} (\bibinfo {year} {2020})}\BibitemShut {NoStop}%
\bibitem [{\citenamefont {Darmawan}\ and\ \citenamefont {Poulin}(2017)}]{tensornet_decoder}%
  \BibitemOpen
  \bibfield  {author} {\bibinfo {author} {\bibfnamefont {Andrew~S}\ \bibnamefont {Darmawan}}\ and\ \bibinfo {author} {\bibfnamefont {David}\ \bibnamefont {Poulin}},\ }\bibfield  {title} {\enquote {\bibinfo {title} {Tensor-network simulations of the surface code under realistic noise},}\ }\href@noop {} {\bibfield  {journal} {\bibinfo  {journal} {Physical review letters}\ }\textbf {\bibinfo {volume} {119}},\ \bibinfo {pages} {040502} (\bibinfo {year} {2017})}\BibitemShut {NoStop}%
\bibitem [{\citenamefont {Bombin}\ and\ \citenamefont {Martin-Delgado}(2006)}]{color_code}%
  \BibitemOpen
  \bibfield  {author} {\bibinfo {author} {\bibfnamefont {Hector}\ \bibnamefont {Bombin}}\ and\ \bibinfo {author} {\bibfnamefont {Miguel~Angel}\ \bibnamefont {Martin-Delgado}},\ }\bibfield  {title} {\enquote {\bibinfo {title} {Topological quantum distillation},}\ }\href@noop {} {\bibfield  {journal} {\bibinfo  {journal} {Physical review letters}\ }\textbf {\bibinfo {volume} {97}},\ \bibinfo {pages} {180501} (\bibinfo {year} {2006})}\BibitemShut {NoStop}%
\bibitem [{\citenamefont {Takada}\ \emph {et~al.}(2023)\citenamefont {Takada}, \citenamefont {Takeuchi},\ and\ \citenamefont {Fujii}}]{takada}%
  \BibitemOpen
  \bibfield  {author} {\bibinfo {author} {\bibfnamefont {Yugo}\ \bibnamefont {Takada}}, \bibinfo {author} {\bibfnamefont {Yusaku}\ \bibnamefont {Takeuchi}}, \ and\ \bibinfo {author} {\bibfnamefont {Keisuke}\ \bibnamefont {Fujii}},\ }\bibfield  {title} {\enquote {\bibinfo {title} {Highly accurate decoder for topological color codes with simulated annealing},}\ }\href@noop {} {\  (\bibinfo {year} {2023})},\ \Eprint {http://arxiv.org/abs/2303.01348} {arXiv:2303.01348 [quant-ph]} \BibitemShut {NoStop}%
\bibitem [{\citenamefont {Couvreur}\ \emph {et~al.}(2013)\citenamefont {Couvreur}, \citenamefont {Delfosse},\ and\ \citenamefont {Z{\'e}mor}}]{Delfosse}%
  \BibitemOpen
  \bibfield  {author} {\bibinfo {author} {\bibfnamefont {Alain}\ \bibnamefont {Couvreur}}, \bibinfo {author} {\bibfnamefont {Nicolas}\ \bibnamefont {Delfosse}}, \ and\ \bibinfo {author} {\bibfnamefont {Gilles}\ \bibnamefont {Z{\'e}mor}},\ }\bibfield  {title} {\enquote {\bibinfo {title} {A construction of quantum ldpc codes from cayley graphs},}\ }\href@noop {} {\bibfield  {journal} {\bibinfo  {journal} {IEEE transactions on information theory}\ }\textbf {\bibinfo {volume} {59}},\ \bibinfo {pages} {6087--6098} (\bibinfo {year} {2013})}\BibitemShut {NoStop}%
\bibitem [{\citenamefont {Chen}\ \emph {et~al.}(2005)\citenamefont {Chen}, \citenamefont {Dholakia}, \citenamefont {Eleftheriou}, \citenamefont {Fossorier},\ and\ \citenamefont {Hu}}]{LDPC_IBM}%
  \BibitemOpen
  \bibfield  {author} {\bibinfo {author} {\bibfnamefont {Jinghu}\ \bibnamefont {Chen}}, \bibinfo {author} {\bibfnamefont {Ajay}\ \bibnamefont {Dholakia}}, \bibinfo {author} {\bibfnamefont {Evangelos}\ \bibnamefont {Eleftheriou}}, \bibinfo {author} {\bibfnamefont {Marc~PC}\ \bibnamefont {Fossorier}}, \ and\ \bibinfo {author} {\bibfnamefont {Xiao-Yu}\ \bibnamefont {Hu}},\ }\bibfield  {title} {\enquote {\bibinfo {title} {Reduced-complexity decoding of ldpc codes},}\ }\href@noop {} {\bibfield  {journal} {\bibinfo  {journal} {IEEE transactions on communications}\ }\textbf {\bibinfo {volume} {53}},\ \bibinfo {pages} {1288--1299} (\bibinfo {year} {2005})}\BibitemShut {NoStop}%
\bibitem [{\citenamefont {Xu}\ \emph {et~al.}(2023)\citenamefont {Xu}, \citenamefont {Ataides}, \citenamefont {Pattison}, \citenamefont {Raveendran}, \citenamefont {Bluvstein}, \citenamefont {Wurtz}, \citenamefont {Vasic}, \citenamefont {Lukin}, \citenamefont {Jiang},\ and\ \citenamefont {Zhou}}]{LDPC_Lukin}%
  \BibitemOpen
  \bibfield  {author} {\bibinfo {author} {\bibfnamefont {Qian}\ \bibnamefont {Xu}}, \bibinfo {author} {\bibfnamefont {J}~\bibnamefont {Ataides}}, \bibinfo {author} {\bibfnamefont {Christopher~A}\ \bibnamefont {Pattison}}, \bibinfo {author} {\bibfnamefont {Nithin}\ \bibnamefont {Raveendran}}, \bibinfo {author} {\bibfnamefont {Dolev}\ \bibnamefont {Bluvstein}}, \bibinfo {author} {\bibfnamefont {Jonathan}\ \bibnamefont {Wurtz}}, \bibinfo {author} {\bibfnamefont {Bane}\ \bibnamefont {Vasic}}, \bibinfo {author} {\bibfnamefont {Mikhail~D}\ \bibnamefont {Lukin}}, \bibinfo {author} {\bibfnamefont {Liang}\ \bibnamefont {Jiang}}, \ and\ \bibinfo {author} {\bibfnamefont {Hengyun}\ \bibnamefont {Zhou}},\ }\bibfield  {title} {\enquote {\bibinfo {title} {Constant-overhead fault-tolerant quantum computation with reconfigurable atom arrays},}\ }\href@noop {} {\bibfield  {journal} {\bibinfo  {journal} {arXiv preprint arXiv:2308.08648}\ } (\bibinfo {year} {2023})}\BibitemShut {NoStop}%
\bibitem [{\citenamefont {Fujii}\ \emph {et~al.}(2014)\citenamefont {Fujii}, \citenamefont {Negoro}, \citenamefont {Imoto},\ and\ \citenamefont {Kitagawa}}]{fujii_prx}%
  \BibitemOpen
  \bibfield  {author} {\bibinfo {author} {\bibfnamefont {Keisuke}\ \bibnamefont {Fujii}}, \bibinfo {author} {\bibfnamefont {Makoto}\ \bibnamefont {Negoro}}, \bibinfo {author} {\bibfnamefont {Nobuyuki}\ \bibnamefont {Imoto}}, \ and\ \bibinfo {author} {\bibfnamefont {Masahiro}\ \bibnamefont {Kitagawa}},\ }\bibfield  {title} {\enquote {\bibinfo {title} {Measurement-free topological protection using dissipative feedback},}\ }\href@noop {} {\bibfield  {journal} {\bibinfo  {journal} {Physical Review X}\ }\textbf {\bibinfo {volume} {4}},\ \bibinfo {pages} {041039} (\bibinfo {year} {2014})}\BibitemShut {NoStop}%
\bibitem [{\citenamefont {Fujisaki}\ \emph {et~al.}(2022)\citenamefont {Fujisaki}, \citenamefont {Oshima}, \citenamefont {Sato},\ and\ \citenamefont {Fujii}}]{fujisaki_1}%
  \BibitemOpen
  \bibfield  {author} {\bibinfo {author} {\bibfnamefont {Jun}\ \bibnamefont {Fujisaki}}, \bibinfo {author} {\bibfnamefont {Hirotaka}\ \bibnamefont {Oshima}}, \bibinfo {author} {\bibfnamefont {Shintaro}\ \bibnamefont {Sato}}, \ and\ \bibinfo {author} {\bibfnamefont {Keisuke}\ \bibnamefont {Fujii}},\ }\bibfield  {title} {\enquote {\bibinfo {title} {Practical and scalable decoder for topological quantum error correction with an ising machine},}\ }\href@noop {} {\bibfield  {journal} {\bibinfo  {journal} {Physical Review Research}\ }\textbf {\bibinfo {volume} {4}},\ \bibinfo {pages} {043086} (\bibinfo {year} {2022})}\BibitemShut {NoStop}%
\bibitem [{\citenamefont {Fujisaki}\ \emph {et~al.}(2023)\citenamefont {Fujisaki}, \citenamefont {Maruyama}, \citenamefont {Oshima}, \citenamefont {Sato}, \citenamefont {Sakashita}, \citenamefont {Takeuchi},\ and\ \citenamefont {Fujii}}]{fujisaki_2}%
  \BibitemOpen
  \bibfield  {author} {\bibinfo {author} {\bibfnamefont {Jun}\ \bibnamefont {Fujisaki}}, \bibinfo {author} {\bibfnamefont {Kazunori}\ \bibnamefont {Maruyama}}, \bibinfo {author} {\bibfnamefont {Hirotaka}\ \bibnamefont {Oshima}}, \bibinfo {author} {\bibfnamefont {Shintaro}\ \bibnamefont {Sato}}, \bibinfo {author} {\bibfnamefont {Tatsuya}\ \bibnamefont {Sakashita}}, \bibinfo {author} {\bibfnamefont {Yusaku}\ \bibnamefont {Takeuchi}}, \ and\ \bibinfo {author} {\bibfnamefont {Keisuke}\ \bibnamefont {Fujii}},\ }\bibfield  {title} {\enquote {\bibinfo {title} {Quantum error correction with an ising machine under circuit-level noise},}\ }\href@noop {} {\bibfield  {journal} {\bibinfo  {journal} {arXiv preprint arXiv:2308.00369}\ } (\bibinfo {year} {2023})}\BibitemShut {NoStop}%
\bibitem [{ope()}]{openjij}%
  \BibitemOpen
  \href@noop {} {\enquote {\bibinfo {title} {{OpenJij}},}\ }\bibinfo {note} {\url{https://www.openjij.org/}}\BibitemShut {NoStop}%
\bibitem [{da1()}]{da1}%
  \BibitemOpen
  \href@noop {} {\enquote {\bibinfo {title} {{Official website of Fujitsu's Digital Annealer}},}\ }\bibinfo {howpublished} {\url{https://www.fujitsu.com/global/services/business-services/digital-annealer/}}\BibitemShut {NoStop}%
\bibitem [{\citenamefont {Aramon}\ \emph {et~al.}(2019)\citenamefont {Aramon}, \citenamefont {Rosenberg}, \citenamefont {Valiante}, \citenamefont {Miyazawa}, \citenamefont {Tamura},\ and\ \citenamefont {Katzgraber}}]{da2}%
  \BibitemOpen
  \bibfield  {author} {\bibinfo {author} {\bibfnamefont {M.}~\bibnamefont {Aramon}}, \bibinfo {author} {\bibfnamefont {G.}~\bibnamefont {Rosenberg}}, \bibinfo {author} {\bibfnamefont {E.}~\bibnamefont {Valiante}}, \bibinfo {author} {\bibfnamefont {T.}~\bibnamefont {Miyazawa}}, \bibinfo {author} {\bibfnamefont {H.}~\bibnamefont {Tamura}}, \ and\ \bibinfo {author} {\bibfnamefont {H.~G.}\ \bibnamefont {Katzgraber}},\ }\bibfield  {title} {\enquote {\bibinfo {title} {Physics-inspired optimization for quad-ratic unconstrained problems using a digital annealer},}\ }\href@noop {} {\bibfield  {journal} {\bibinfo  {journal} {Frontiers in Physics}\ }\textbf {\bibinfo {volume} {7}} (\bibinfo {year} {2019})}\BibitemShut {NoStop}%
\bibitem [{\citenamefont {Sao}\ \emph {et~al.}(2019)\citenamefont {Sao}, \citenamefont {Watanabe}, \citenamefont {Musha},\ and\ \citenamefont {Utsunomiya}}]{da3}%
  \BibitemOpen
  \bibfield  {author} {\bibinfo {author} {\bibfnamefont {M.}~\bibnamefont {Sao}}, \bibinfo {author} {\bibfnamefont {H.}~\bibnamefont {Watanabe}}, \bibinfo {author} {\bibfnamefont {Y.}~\bibnamefont {Musha}}, \ and\ \bibinfo {author} {\bibfnamefont {A.}~\bibnamefont {Utsunomiya}},\ }\bibfield  {title} {\enquote {\bibinfo {title} {Application of digital annealer for faster combinatorial optimization},}\ }\href@noop {} {\bibfield  {journal} {\bibinfo  {journal} {FUJITSU SCIENTIFIC \& TECHNICAL JOURNAL}\ }\textbf {\bibinfo {volume} {55}},\ \bibinfo {pages} {45} (\bibinfo {year} {2019})}\BibitemShut {NoStop}%
\bibitem [{\citenamefont {Matsubara}\ \emph {et~al.}(2020)\citenamefont {Matsubara}, \citenamefont {Takatsu}, \citenamefont {Miyazawa}, \citenamefont {Shibasaki}, \citenamefont {Watanabe}, \citenamefont {Takemoto},\ and\ \citenamefont {Tamura}}]{da4}%
  \BibitemOpen
  \bibfield  {author} {\bibinfo {author} {\bibfnamefont {S.}~\bibnamefont {Matsubara}}, \bibinfo {author} {\bibfnamefont {M.}~\bibnamefont {Takatsu}}, \bibinfo {author} {\bibfnamefont {T.}~\bibnamefont {Miyazawa}}, \bibinfo {author} {\bibfnamefont {T.}~\bibnamefont {Shibasaki}}, \bibinfo {author} {\bibfnamefont {Y.}~\bibnamefont {Watanabe}}, \bibinfo {author} {\bibfnamefont {K.}~\bibnamefont {Takemoto}}, \ and\ \bibinfo {author} {\bibfnamefont {H.}~\bibnamefont {Tamura}},\ }\bibfield  {title} {\enquote {\bibinfo {title} {Digital annealer for high-speed solving of combinatorial optimization problems and its applications},}\ }\href@noop {} {\bibfield  {journal} {\bibinfo  {journal} {25th Asia and South Pacific Design Automation Conference (ASP-DAC)}\ } (\bibinfo {year} {2020})}\BibitemShut {NoStop}%
\bibitem [{cpl()}]{cplex}%
  \BibitemOpen
  \href@noop {} {\enquote {\bibinfo {title} {{IBM ILOG CPLEX Optimizer}},}\ }\bibinfo {note} {\url{https://www.ibm.com/products/ilog-cplex optimization-studio/cplex-optimizer}}\BibitemShut {NoStop}%
\bibitem [{\citenamefont {Fowler}\ \emph {et~al.}(2009)\citenamefont {Fowler}, \citenamefont {Stephens},\ and\ \citenamefont {Groszkowski}}]{surface1}%
  \BibitemOpen
  \bibfield  {author} {\bibinfo {author} {\bibfnamefont {Austin~G}\ \bibnamefont {Fowler}}, \bibinfo {author} {\bibfnamefont {Ashley~M}\ \bibnamefont {Stephens}}, \ and\ \bibinfo {author} {\bibfnamefont {Peter}\ \bibnamefont {Groszkowski}},\ }\bibfield  {title} {\enquote {\bibinfo {title} {High-threshold universal quantum computation on the surface code},}\ }\href@noop {} {\bibfield  {journal} {\bibinfo  {journal} {Physical Review A}\ }\textbf {\bibinfo {volume} {80}},\ \bibinfo {pages} {052312} (\bibinfo {year} {2009})}\BibitemShut {NoStop}%
\bibitem [{\citenamefont {Kitaev}(2003)}]{surface2}%
  \BibitemOpen
  \bibfield  {author} {\bibinfo {author} {\bibfnamefont {A~Yu}\ \bibnamefont {Kitaev}},\ }\bibfield  {title} {\enquote {\bibinfo {title} {Fault-tolerant quantum computation by anyons},}\ }\href@noop {} {\bibfield  {journal} {\bibinfo  {journal} {Annals of physics}\ }\textbf {\bibinfo {volume} {303}},\ \bibinfo {pages} {2--30} (\bibinfo {year} {2003})}\BibitemShut {NoStop}%
\bibitem [{\citenamefont {Fujii}\ and\ \citenamefont {Hayashi}(2017)}]{fujii2017}%
  \BibitemOpen
  \bibfield  {author} {\bibinfo {author} {\bibfnamefont {Keisuke}\ \bibnamefont {Fujii}}\ and\ \bibinfo {author} {\bibfnamefont {Masahito}\ \bibnamefont {Hayashi}},\ }\bibfield  {title} {\enquote {\bibinfo {title} {Verifiable fault tolerance in measurement-based quantum computation},}\ }\href@noop {} {\bibfield  {journal} {\bibinfo  {journal} {Physical Review A}\ }\textbf {\bibinfo {volume} {96}},\ \bibinfo {pages} {030301} (\bibinfo {year} {2017})}\BibitemShut {NoStop}%
\bibitem [{\citenamefont {Poulin}(2006)}]{Poulin}%
  \BibitemOpen
  \bibfield  {author} {\bibinfo {author} {\bibfnamefont {David}\ \bibnamefont {Poulin}},\ }\bibfield  {title} {\enquote {\bibinfo {title} {Optimal and efficient decoding of concatenated quantum block codes},}\ }\href@noop {} {\bibfield  {journal} {\bibinfo  {journal} {Physical Review A}\ }\textbf {\bibinfo {volume} {74}},\ \bibinfo {pages} {052333} (\bibinfo {year} {2006})}\BibitemShut {NoStop}%
\bibitem [{\citenamefont {Kowalsky}\ \emph {et~al.}(2022)\citenamefont {Kowalsky}, \citenamefont {Albash}, \citenamefont {Hen},\ and\ \citenamefont {Lidar}}]{da_annealing_time}%
  \BibitemOpen
  \bibfield  {author} {\bibinfo {author} {\bibfnamefont {Matthew}\ \bibnamefont {Kowalsky}}, \bibinfo {author} {\bibfnamefont {Tameem}\ \bibnamefont {Albash}}, \bibinfo {author} {\bibfnamefont {Itay}\ \bibnamefont {Hen}}, \ and\ \bibinfo {author} {\bibfnamefont {Daniel~A}\ \bibnamefont {Lidar}},\ }\bibfield  {title} {\enquote {\bibinfo {title} {3-regular three-xorsat planted solutions benchmark of classical and quantum heuristic optimizers},}\ }\href@noop {} {\bibfield  {journal} {\bibinfo  {journal} {Quantum Science and Technology}\ }\textbf {\bibinfo {volume} {7}},\ \bibinfo {pages} {025008} (\bibinfo {year} {2022})}\BibitemShut {NoStop}%
\bibitem [{\citenamefont {Yoshimura}\ \emph {et~al.}(2016)\citenamefont {Yoshimura}, \citenamefont {Hayashi}, \citenamefont {Okuyama},\ and\ \citenamefont {Yamaoka}}]{FPGA1}%
  \BibitemOpen
  \bibfield  {author} {\bibinfo {author} {\bibfnamefont {Chihiro}\ \bibnamefont {Yoshimura}}, \bibinfo {author} {\bibfnamefont {Masato}\ \bibnamefont {Hayashi}}, \bibinfo {author} {\bibfnamefont {Takuya}\ \bibnamefont {Okuyama}}, \ and\ \bibinfo {author} {\bibfnamefont {Masanao}\ \bibnamefont {Yamaoka}},\ }\bibfield  {title} {\enquote {\bibinfo {title} {Fpga-based annealing processor for ising model},}\ }in\ \href@noop {} {\emph {\bibinfo {booktitle} {2016 Fourth International Symposium on Computing and Networking (CANDAR)}}}\ (\bibinfo {organization} {IEEE},\ \bibinfo {year} {2016})\ pp.\ \bibinfo {pages} {436--442}\BibitemShut {NoStop}%
\bibitem [{\citenamefont {Patel}\ \emph {et~al.}(2020)\citenamefont {Patel}, \citenamefont {Chen}, \citenamefont {Canoza},\ and\ \citenamefont {Salahuddin}}]{FPGA2}%
  \BibitemOpen
  \bibfield  {author} {\bibinfo {author} {\bibfnamefont {Saavan}\ \bibnamefont {Patel}}, \bibinfo {author} {\bibfnamefont {Lili}\ \bibnamefont {Chen}}, \bibinfo {author} {\bibfnamefont {Philip}\ \bibnamefont {Canoza}}, \ and\ \bibinfo {author} {\bibfnamefont {Sayeef}\ \bibnamefont {Salahuddin}},\ }\bibfield  {title} {\enquote {\bibinfo {title} {Ising model optimization problems on a fpga accelerated restricted boltzmann machine},}\ }\href@noop {} {\bibfield  {journal} {\bibinfo  {journal} {arXiv preprint arXiv:2008.04436}\ } (\bibinfo {year} {2020})}\BibitemShut {NoStop}%
\end{thebibliography}%

\end{document}